\documentclass[twocolumn,superscriptaddress,showpacs,amsmath,amssymb,aip,reprint,jcp]{revtex4-1}
\usepackage[pdftex]{graphicx}
\usepackage[]{subfigure}
\usepackage{hyperref}
\newcommand{\bi}{\begin{itemize}}
\newcommand{\ei}{\end{itemize}}
\newcommand{\benum}{\begin{enumerate}}
\newcommand{\eenum}{\end{enumerate}}
\newcommand{\be}{\begin{equation}}
\newcommand{\ee}{\end{equation}}
\newcommand{\bea}{\begin{eqnarray}}
\newcommand{\eea}{\end{eqnarray}}
\newcommand{\beas}{\begin{eqnarray*}}
\newcommand{\eeas}{\end{eqnarray*}}
\newcommand{\bfr}{{\bf r}}

\newcommand{\CtwoHtwoOtwo}{\mathrm{C_2H_2O_2}}
\newcommand{\Stwo}{\mathrm{S_2}}
\newcommand{\CthreeHfour}{\mathrm{C_3H_4}}
\newcommand{\CfourHeight}{\mathrm{C_4H_8}}
\newcommand{\SiHfour}{\mathrm{SiH_4}}

\newcommand{\equal}{\!=\!}
\newcommand{\lapln}{\nabla^2n}
\newcommand{\gradn}{\nabla n}
\newcommand{\gradnsq}{ \left|\nabla n\right|^2}
\newcommand{\gradnsqn}{ \left|\nabla n\right|^2/n}
\newcommand{\barlapln}{\overline{\nabla^2n}}

\newcommand{\tauTF}{\tau_{TF}}
\newcommand{\tauvW}{\tau_{vW}}
\newcommand{\uvW}{u_{vW}}
\newcommand{\tauGEA}{\tau_{GEA}}

\newcommand{\taumGGA}{\tau_{mGGA}}
\newcommand{\tauKS}{\tau_{KS}}
\newcommand{\TKS}{T_{KS}}

\newcommand{\TITLE}{Visualization and analysis of the Kohn-Sham kinetic energy density and its orbital-free description in molecules} 

\begin{document}

\preprint{Cancio, Stewart and Kuna, preprint 2016}

\title{ \TITLE }
\author{Antonio C. Cancio}
\author{Dane Stewart}
\author{Aeryk Kuna}
\affiliation{Department of Physics and Astronomy, Ball State University, Muncie, Indiana 47306}
\email[]{accancio@bsu.edu}

%aeryk.kuna@gmail.com  Aeryk I. Kuna
%destewart@bsu.edu Dane E. Stewart

\date{\today}

\begin{abstract} 
We visualize the Kohn-Sham kinetic energy density (KED), and the ingredients -- the electron density, its gradient and Laplacian -- used to construct orbital-free models of it, for the AE6 test set of molecules.  These are compared to related quantities used in metaGGA's, to characterize two important limits -- the gradient expansion and the localized-electron limit typified by the covalent bond.  We find the second-order gradient expansion of the KED to be a surprisingly successful predictor of the exact KED, particularly at low densities where this approximation fails for exchange.  This contradicts the conjointness conjecture that the optimal enhancement factors for orbital-free kinetic and exchange energy functionals are closely similar in form.  In addition we find significant problems with a recent metaGGA-level orbital-free KED, especially for regions of strong electron localization.  We define an orbital-free description of electron localization and a revised metaGGA that improves upon atomization energies significantly.
\end{abstract}

\keywords{Density Functional Theory, Laplacian of the Density, Gradient of the Density, Kinetic Energy Density, metaGGA, orbital-free DFT, AE6, ELF}

\maketitle

\section{Introduction}

The Kohn-Sham kinetic energy
density (KED) -- the kinetic energy per volume defined by the orbitals
generated by the Kohn-Sham equation --  
plays a central role in the development of density functional theory (DFT).
In the ``Jacob's Ladder" paradigm for 
characterizing the exchange-correlation (XC) energy in 
density functional theory,~\cite{JacobsLadder}
the KED is the key variable of the central, metaGGA rung of 
functionals.~\cite{BeckemGGA,PKZB,TPSS,ML06}
As a local energy density, it 
provides information about electronic structure 
complementary to that provided by the local electron density and 
its derivatives that describe lower rungs of DFT. 
Particularly important is its ability to 
distinguish between regions of electron 
localization,\cite{beckeELF, MGGA-MS, FAM14} for which 
self-interaction error is important, 
and regions of delocalization such as metals where they are not.

The centrality of the KED in DFT development is highlighted by the 
implicit role it plays in the rungs of DFT lower than the metaGGA.  These 
may be thought of as %a ladder of approximations to the KED 
a Jacob's ladder of approximations to the KED
as much as one of approximations to the exchange-correlation energy.
The lowest rung of the XC ladder, the local density approximation or 
LDA,~\cite{KohnSham}
corresponds to the Thomas-Fermi approximation~\cite{Thomas27,Fermi28} 
to the KED.
The more commonly used generalized gradient approximation (GGA) 
introduces, in addition to the local density, the gradient of the density 
as a variable in XC functional construction.  
The same information is contained in the von~Weizs\"acker 
KED~\cite{vonWeizsacker} 
that describes the KE of localized electron pairs. 
It has been used to generate a large number of GGA's for the KED, 
both empirical~\cite{KJTH09,TranWesolowski,LacksGordon94,Thakkar92} and 
nonempirical,~\cite{APBE,KCST13}
though not with the success they have enjoyed in describing the XC energy.
To describe the XC energy at the next, metaGGA, level of the theory, 
not only the KED, but also
the Laplacian of the 
density~\cite{JemmerKnowles, CancioExlapl, Pittalis, PittalisAlso}  
may be used as an additional variable in functional construction.
A metaGGA description of the KED is thus possible, using the 
Laplacian.\cite{PerdewConstantin,KJTH09,LCFSLapl}
The similarity of the KED and XC functional ladders 
leads to a conjointness conjecture\cite{LeeLeeParr91}
that the optimal orbital-free correction to the
Thomas-Fermi KE is similar in form to that for LDA exchange. 
More importantly, it enables one to apply lessons learned in 
constructing the one functional to constructing the other.  
This is important for orbital-free DFT, in which the Kohn-Sham KED 
is replaced by an explicit functional of the density, removing completely 
the need for orbitals.

The orbital-free modeling of the KED 
has taken on increasing importance in recent years.~\cite{KCT2013,AkimovPrezhdo}
Given a cubic scaling in the number orbitals, the Kohn-Sham method becomes 
prohibitively expensive
for large-scale applications that require the accuracy of atomistic simulation.
These involve applications such as the dynamics 
of nanoscale materials~\cite{AkimovPrezhdo} 
requiring intrinsically large system size,
high throughput as in alloy design, or a need for large number of 
excited states, as can occur for finite-temperature applications
such as warm dense matter (WDM).~\cite{WDM,KarasievWDM}  %WDMBasicNeeds
A completely orbital-free density functional theory (OFDFT), using an 
orbital-free expression for the  kinetic energy, becomes an important
tool in these cases.  
Unfortunately, OFDFT is inherently less accurate than Kohn-Sham DFT; 
for example, the Thomas-Fermi approximation
is unable to predict molecular binding,~\cite{Teller} something the LDA
has no problem doing.  
Nevertheless, nonlocal models~\cite{WGC1999, ErratumWGC1999,
HuangCarter10,KLXC14}
have achieved reasonably high accuracy, 
allowing for impressive calculations
for solid-state applications,~\cite{HungCarter11, ShinCarter13} albeit within the limitation of requiring
different functionals for different 
material classes.
And, with a focus on improving potentials and thus forces in the context of 
WDM, a number of GGA-level OFDFT's have been developed in recent 
years.~\cite{KJTH09,KCST13,KCT2013}
These coincide with improvements in infrastructure for 
practical calculations.~\cite{KarasievTrickey12,PROFESS}

A third role of the KED has been 
as a tool for visualizing the electronic structure of the chemical bond.
The kinetic energy density has been the subject of 
investigation~\cite{RS2009,Tachibana2001}
particularly as a localized-orbital 
locator (LOL),~\cite{SchmiderBecke,JacobsenPCCP2013}
and an impressive number of related quantities have been defined and 
investigated as well.~\cite{AAH2010}
Perhaps the most popular is the electron localization factor ELF, which is 
based upon a comparison of the Kohn-Sham to Thomas-Fermi and von~Weizs\"acker 
KED's.~\cite{beckeELF,SavinELF,BurnusMarquesGross} It is of particular 
importance for the development of metaGGA's for the XC energy and in the 
conceptual understanding of why they work.~\cite{FAM14,MGGA-MS}
Also of note is the quantum theory of atoms in 
molecules (QTAIM),\cite{QTAIM,BaderReview,MMK2011}
an approach to visualization which is in 
some ways an orbital-free version of ELF analysis, using gradients and 
Laplacians of the density to analyze bonding structures. 

Despite the strong connection between the arguments used to build
XC metaGGA's~\cite{BeckemGGA,TPSS,revTPSS,MGGA-MS} and those used to
visualize the chemical bond, the tool of visualization
has not often been used to provide feedback into DFT development.
The properties of the exchange-correlation hole, describing the hole
around an electron caused by Pauli exclusion and Coulomb repulsion have
been an important tool in the construction of both GGA's and their 
successors.~\cite{PBW_GGA, Beckenxc, BCGPacGGA}              %%%PBE-ontop also, 
In particular, the visualization of the hole has been a valuable tool in 
assessing the accuracy of DFT's.~\cite{Hood1, Cancio2RowPaper}
However, the exchange-correlation hole is a difficult many-body
calculation, and the dependence of measurables like 
the atomization energy or bond length on the nature of the
XC hole occurs implicitly through the mediation of 
complex functionals and thus is hard to determine.  (But
connections can sometimes be made.~\cite{CancioChouHood,CancioLapl})
In the case of the KED, however, visualization is of 
direct help~\cite{XC15,XC15comment,XC15reply} -- how an orbital-free density 
functional theory 
for the Kohn-Sham KED actually compares to the real thing requires no more 
than running a standard DFT code and visualizing the results.

In this paper, we perform highly converged Kohn-Sham DFT 
calculations and visualize the electron density,
its gradient and Laplacian, the KED and some approximations
for these used in DFT, for the AE6 test set of molecules, in a pseudopotential
plane wave approach.  

The AE6 test set~\cite{AE6} is a set of 6 molecules --
Cyclobutane ($\CfourHeight$), Propyne ($\CthreeHfour$), 
Glyoxal ($\CtwoHtwoOtwo$), Silicon Monoxide (SiO), 
Disulfur (S$_2$), and Silicon Tetrahydride ($\SiHfour$)
-- 
chosen for their ability to reproduce the average atomization energy 
of common DFT's over much larger test sets. 
For such a small set the AE6 shows a richness of bonding scenarios --
single, double, and triple bonds, 
covalent to nearly ionic, including first and second row atoms, 
and a large-cation, small-anion system similar to important semiconductors 
like GaN.  Thus it
covers many situations commonly seen in organic chemistry and in 
semiconductors as well.  

Our motive for using pseudopotentials is two-fold.
First of all, many current OFDFT applications rely upon the use of 
pseudopotentials,~\cite{HuangCarter10,KLXC14} although
more accurate approaches do exist.~\cite{ofdftgpaw}
More importantly, the pseudopotential plane-wave approach
%while it introduces errors into the chemical 
%characterization of the test set, %this is not the focus of the current study.
permits an arbitrary convergence of the particle density associated with
the pseudopotential and thus a map between a $v-$representable density and 
the related KE density that is as accurate as possible.  It thus gives
insight into the universal map between kinetic energy and density that
is a corollary of the Hohenberg-Kohn theorem.
Although the method does not produce the correct density for 
real molecules, and thus introduces errors into the chemical 
characterization of the test set, it arguably gives us simpler problem to 
model, and much of what is learned for pseudopotential systems should help to 
construct functionals for the all-electron case.~\cite{KT15footnote}  
The use of pseudopotentials enables particularly
the study of asymptotic features not possible with a typical 
gaussian basis set.
%although questions have been raised as to their reliability~\cite{???} and 

%%%% MAYBE --  Note

Finally, the choice of exchange-correlation functional is irrelevant to the
universal mapping between the Kohn-Sham \textit{kinetic} energy and the 
charge density, in which the electrostatic potential energy plays no role.
We work with the LDA and PBE exchange-correlation functionals, which
produce reasonably accurate bond lengths for the test set and should produce
densities and orbitals close to the exact ones for pseudopotential systems.  

In our visualization, we have deemphasized (but do not ignore) the 
ELF, already studied extensively for a large number of molecular systems. 
We look rather at the basic ingredients of the orbital-free KED, the electron 
density $n$, and related derivatives $\gradnsqn$ and $\lapln$, 
focusing especially on applications of their use in DFT. 
One is a common approximation based upon the gradient expansion
in the limit of slowly varying densities used in many metaGGA's to replace 
$\lapln$, a natural descriptor in this limit, for $\tau_{KS}$. 
The second is a sophisticated metaGGA-level orbital-free model of 
the Kohn-Sham KED, the mGGA.~\cite{PerdewConstantin}
This takes advantage of lessons learned in developing metaGGA's for 
exchange, particularly of defining and respecting key constraints and 
limiting cases for the kinetic energy.  Despite the promise of its
design philosophy, the mGGA has deficiencies -- its potential
does not bind molecules~\cite{KJTH09} and even used non-self-consistently
fails to improve upon Thomas-Fermi predictions of atomization 
energies.\cite{LCFSLapl}
However, it is of value as a starting point of thinking how to construct
a metaGGA; and since it is a model of the kinetic energy 
density as such, it is directly testable by visualization of this quantity.
Our investigation of the mGGA shows, despite its
excellent description of atomic KED's, surprising
failures in its description of the KED of bonds, and thus
in its prediction of atomization energies.
Our visualization work makes it easy to diagnose and suggest a fix to 
this problem, one which defines, and demonstrates at least 
in an \textit{ad hoc} fashion, a potential lower bound to the KED. 

The rest of this paper is organized as follows: 
Sec.~\ref{sec:theory} 
describes the theoretical background of the paper -- the density functional 
theory of the kinetic energy and its relation to exchange in metaGGA's. 
Sec.~\ref{sec:methodology} covers the basic methodology used.  
Sec.~\ref{sec:results} 
details the chief results of visualization, while
Sec.~\ref{sec:discussion} 
applies the lessons learned to construct and make preliminary tests of
 a correction to the Perdew-Constantin mGGA
and Sec.~\ref{sec:conclusion} 
presents our conclusions.

\section{Theory \label{sec:theory}}
The positive definite form of the kinetic energy density in Kohn-Sham theory 
is given by  
\be
    \tauKS = \frac{1}{2} \sum_i^{occup} f_i \left| \nabla \phi_i \right|^2,
    \label{eq:tauks}
\ee 
where $\phi_i$ are Kohn-Sham orbitals from which the electron density 
is constructed:
\be
      n = \sum_i^{occup} f_i \left| \phi_i \right|^2,
     \label{eq:dens}
\ee
and $f_i$ is the occupation number of each orbital.
Integration over all space gives the kinetic energy
\be
     \TKS[n] = \int \tauKS(\bfr) d^3r.
\ee
A generalization in terms of the spin density and spin-decomposed
KED's is easily constructed by restricting the
sums in the equations above to a specific spin species but will not be
considered here.  
The KED is well defined only up to the arbitrary addition of
a divergence of a vector function -- the integration of such an addition is
zero and leaves the physical measurable $\TKS$ unchanged.
Thus any number of physically equivalent KED's may be constructed, with 
a common alternative to Eq.~(\ref{eq:tauks}) being
\be
    \tauKS' = -\frac{1}{2} \sum_i^{occup} f_i \phi_i^* \nabla^2 \phi_i = 
              \tauKS - \frac{1}{4}\nabla^2 n.
    \label{eq:tauksp}
\ee
The value of Eq.~(\ref{eq:tauks}) is that it is 
positive-definite like the particle density, and that a number of 
properties of the 
KE are conveniently framed in terms of it. %this choice of energy density.

The key principle for this paper is that since $\TKS[n]$ is a functional of the 
ground state electron density $n$, $\tauKS$ must be one too. There exists
some map $\tauKS[n]$ from Eq.~(\ref{eq:dens}) to Eq.~(\ref{eq:tauks}) that 
need not explicitly rely on orbitals.  However the form of this map is unknown,
and unlike the exchange-correlation functional of standard Kohn-Sham theory,
approximate functionals are often far from satisfactory.
Specifically, as is done in the lower rungs of the XC ladder of approximations,
we can define a ``semilocal" model of $\TKS[n]$, 
in terms of functions of the local density and its derivatives:
\be
         \TKS^{approx}[n] = \int \tau^{approx}[n(\bfr), \gradn(\bfr), \lapln(\bfr)] d^3r
    \label{eq:semilocal}
\ee
This is the take-off point for many orbital-free functionals for 
$\TKS$,~\cite{LCFSLapl,KCST13,KJTH09,TranWesolowski,LacksGordon94,Thakkar92,PerdewConstantin}
and the point of view considered in this paper.
At the same time, nonlocal functionals~\cite{JG,WangTeter92,WGC1999,ErratumWGC1999,HuangCarter10} 
take the form
\be
   \TKS[n] = \int\int  n^\alpha(\bfr) W(\bfr,\bfr') n^\beta(\bfr') d^3r d^3r'
\ee
which may be related to the semilocal picture through an expansion 
of the kernel $W(\bfr,\bfr')$ for small $\bfr-\bfr'$.~\cite{HK} 

The lowest level of semilocal functional -- the equivalent to the LDA in
XC functionals -- is the Thomas-Fermi model,
\be
   \tauTF = \frac{3}{10} k_F^2 n\!\sim\!n^{5/3},
\ee
with $k_F = (3 \pi^2 n)^{1/3}$ the fermi wavevector of the homogeneous
electron gas (HEG).  
At the next level of approximation, the gradient 
expansion approximation (GEA)~\cite{Kirzhnits57,BrackJenningsChu} of the KED 
is given by:
\be
      \tauGEA = \tauTF + \frac{1}{72}|\nabla n|^2/n + \frac{1}{6}\nabla^2 n 
                    + O(\nabla^4).
      \label{eq:taugea} 
\ee
Terms up to fourth~\cite{Hodges} and sixth order~\cite{Murphy81} in this 
expansion are known.

As is the case with exchange, 
in order to preserve the proper scaling of $\TKS$ under the uniform scaling of 
the charge density, the form of an orbital-free functional
for the KED is restricted to that of 
a function of scale-invariant quantities  %(such as $p$ and $q$) 
times the local density approximation. 
Thus the GEA can be recast as 
\be
       \tauGEA = \left[1 + \frac{5}{27}p + \frac{20}{9}q\right] \tauTF,
      \label{eq:taugeaalso}
\ee
in terms of invariant quantities: 
\bea
  p &=&  \frac{\gradnsq}{4 k_F^2 n}, \\
  q &=& \frac{\lapln}{4 k_F^2 n}.
   \label{eq:q}
\eea
Similarly, the most general form for a semilocal functional 
is a generalization of the GEA form in terms 
of an enhancement factor $F_S$ modifying $\tauTF$:
\be
       \tau_{semilocal} \!= \! F_S(p,q) \tauTF.
      \label{eq:taugga}
\ee
The enhancement factor $F_S$ for the kinetic energy plays a similar role 
to that for exchange, $F_X$, in conventional GGA's, where
the exchange energy density is expressed as a correction to the LDA
in the form $F_X e_X^{LDA}$. 

In constructing generalized gradient functionals,
it is conventional to omit the term proportional to $\lapln$
in the GEA expansion as this integrates to zero and does not contribute
to the overall kinetic energy.~\cite{Kirzhnits57}  Then, by approximating the 
gradient expansion to all orders in the remaining variable $p$, one obtains 
the next natural step, the GGA.~\cite{Thakkar92}  
However, our goal is to 
visualize the local quantity $\tauKS(\bfr)$, and for this purpose,
the $\lapln$ term in its gradient expansion  cannot be ignored.
Moreover, keeping it is necessary to implement local constraints on
orbital-free approximation to $\tauKS$ 
(and thus constraints on $\TKS$) correctly, and we do so in the work that 
follows. 
$\lapln$ is normally considered as 
a higher-order variable whose introduction in a functional defines the next, 
metaGGA, rung of functionals.

Up to this level of approximation, the process of building a 
kinetic energy functional mirrors that for exchange, so that 
the conjointness conjecture has been made~\cite{LeeLeeParr91} 
that the optimal form for each functional at a given level of approximation are
closely related: $F_S\!\sim\!F_X$.
This relationship has never been explicitly defined, but is normally taken
to be that of nearly identical functional forms with 
different constants.~\cite{TranWesolowski,LacksGordon94,Thakkar92}
This strict conjecture has been demonstrated to 
be wrong,~\cite{KarasievConjoint} but a philosophy of conjointness,
using the experience of designing exchange-correlation functionals to inform
the design of KE functionals, is common practice.~\cite{APBE,KCST13,LCFSLapl}

Nevertheless, there are fundamental differences between the two functionals,
particularly in the physics of the
large inhomogeneity limit $p, |q| \!\gg\! 1$.
For the Kohn-Sham KED in real systems, the most crucial issue 
is the limit of a one-particle system or two-particle 
spin-singlet system.
In this case it reduces to the
the von~Weizs\"acker~\cite{vonWeizsacker} functional:
   \be
       \tauvW = \frac{1}{8} \frac{\left| \nabla n \right|^2}{n}.
        \label{eq:tauvw}
   \ee
This is the exact result for a system of $N$ particles
obeying Bose statistics, so that in the ground state they
occupy a single ground state orbital,
        $\phi_0 \equal \sqrt{n}/N.$
The KED needed to create the density $n(\bfr)$ with fermions, that is, the 
energetic cost of Pauli exclusion, 
is given by the difference between the Kohn-Sham and Wigner KED's and 
must be positive definite:~\cite{Herring86}
   \be
    \tau_{Pauli} = \tauKS - \tauvW \ge 0.
   \ee
Notably, this von~Weizs\"acker lower bound is not respected by the GEA.
If we rewrite Eq.~(\ref{eq:tauvw}) in terms of an enhancement
factor, we find $F_S^{vW} = 5p/3$ -- a dependence on $p$ that is 
nine times faster than that of the GEA.  For $q \equal 0$, $\tauGEA$ 
falls below $\tauvW$ for the relatively modest value of $p \equal 27/40$.  
The constraint can be imposed by changing the coefficient in the gradient 
expansion to 5/3, in which case the slowly-varying limit is incorrect.  
In contrast, exchange is constrained by the Lieb-Oxford 
bound~\cite{Lieb} that limits the contribution from 
the low density tail outside the classically
allowed range of electron.  This limit has no intrinsic tie to the 
single-orbital limit and we shall see that the KED behaves very
differently from exchange in this limit.

Recently a metaGGA-level KED functional of the form of Eq.~(\ref{eq:taugga}), 
the Perdew-Constantin mGGA,~\cite{PerdewConstantin} has been developed by
applying lessons learned in constructing constraint-based exchange-correlation 
functionals.  It satisfies the gradient expansion up to fourth-order
in the limit of slowly varying density and 
the von~Weizs\"acker bound 
and other constraints for large values of $p$ and $q$.
The function interpolates 
between the gradient expansion and 
von Weizs\"acker limits using a nonanalytic but smooth
interpolating function that depends 
on an effective localization measure $z \!=\! F_{GE4-M} \!-\! F_{vW}$, 
with $F_{GE4-M}$ a metaGGA designed to be the best-possible 
analytic functional built 
from the starting point of the slowly varying electron gas.  
%%%%% Corrections here? %%%%%%%
This is explicitly a 
model of the kinetic energy \textit{density}, designed to take the 
place of the KED in exchange-correlation metaGGA's, and thus is
meaningfully tested by means of visualization of the KED.  

So far, the Jacob's Ladder of approximations of the Kohn-Sham KED parallels 
the development of exchange functionals.  A divergence now
occurs in that, for exchange and correlation, the KED %$\tauKS$
itself can be used as a variable for building further approximations. 
In the standard approach~\cite{BeckemGGA} to constructing metaGGA's
for exchange,
the Laplacian of the density, which appears irreducibly in fourth and
higher-order terms in the gradient expansion is introduced implicitly 
through the use of the Kohn-Sham KED.
This is achieved by rewriting the gradient expansion
for $\tauKS$, [Eqs.~(\ref{eq:taugea}--\ref{eq:taugeaalso})], 
to construct a replacement for $\lapln$, good to second order in this 
expansion. 
This ``pseudo-Laplacian" is given by:
\be
     \barlapln = 6\left(\tauKS - \tauTF \right) - \frac{1}{12}|\nabla n|^2/n,
     \label{eq:laplgea}
\ee
which then replaces $\lapln$ in the construction of the metaGGA.
$\barlapln$ approaches $\lapln$
in the limit of slowly varying density,    %($p$ and $q$ small), 
deviating from it only where the $\lapln$ gets 
large, such as at the cusp in the electron density at the nucleus.  
It is unknown how well this approximation works in practice 
for features of electronic structure like covalent bonds, which
locally may have small $p$ and $q$ but are part of systems that are far from 
the slowly-varying limit globally.  This quantity can then serve to test 
the quality of the GEA.

Perhaps the most physically significant role played by the KED in 
a metaGGA is as a measure of electron localization.~\cite{BeckemGGA,MGGA-MS} 
This is done by taking
the ratio of the Pauli contribution to the Kohn-Sham
KED to that of the Thomas-Fermi model,
   \be
         %\alpha = \frac{\Delta \tau}{\tau_{HEG}},
         \alpha = \frac{\tauKS-\tauvW}{\tauTF}.
         \label{eq:alpha}
   \ee
In regions where the KE density is determined predominantly
by a single molecular orbital, 
$\tauKS$ approaches $\tauvW$ and $\alpha \!\rightarrow\! 0$.  
This limit describes %regions dominated by 
single covalent bonds and lone pairs, and generally situations in which the 
self-interaction errors in the GGA and LDA are most acute.  
The HEG, and presumably systems formed by metallic bonds, corresponds to 
$\tauKS~\equal~\tauTF$, $\tauvW\!\sim\!0$ and $\alpha\!\sim\!1$.
Between atomic shells and at low density, 
$\alpha \!\gg\! 1$, potentially tending to $\infty$ for an exponentially 
decaying density if $\tau_{Pauli}$ vanishes more slowly than $n^{5/3}$.  
This limit can be used to detect weak bonds such as van der Waals interactions
and define interstitial regions in semiconductor systems. 
The information on local environment %provided by $\alpha$ 
can then be used to customize gradient approximations for specific 
subsystems.~\cite{MGGA-MS}

It is a short step from $\alpha$ to the electron localization factor 
or ELF~\cite{beckeELF} used in the visualization of electronic structure: 
   \be
         ELF = \frac{1}{1 + \alpha^2}.
   \ee
This converts the information contained in $\alpha$  to a function between
one, when $\alpha\! =\! 0$, to zero ($\alpha \!\rightarrow\! \infty$), 
useful for visualization, but less so in functional construction.
The related LOL~\cite{SchmiderBecke} is closer in form to $\barlapln$, 
and is basically the enhancement factor $F_{KS} \!=\! \tauKS/\tauTF$
recast into a convenient form: $LOL = 1/(1 + F_{KS})$.

Finally we note that the $\alpha$ used in defining the ELF
is also the enhancement factor for the Pauli KE: 
$\tau_{Pauli} = \alpha\tauTF$. And thus one can consider the project of 
constructing OFDFT as intimately tied to the project of visualizing
electronic structure -- constructing 
orbital-free models to the ELF and the information on electron localization 
it contains.  This has been 
the perspective of several recent studies of the KED.~\cite{Finzel15,XC15}
% and orbital free models building models to $\alpha$ in
%an orbital free fashion.

\section{Methodology \label{sec:methodology}}
As noted in the introduction, we use the plane-wave pseudopotential
method for performing DFT calculations --  this allows us to solve nearly 
exactly
the Kohn-Sham equation for a model system and acquire highly accurate
orbitals, but for an approximate system. %, focused on bonding features.  
For this purpose, the ABINIT plane-wave pseudopotential code~\cite{Abinit1,Abinit2,Abinit3} was employed with an LDA and PBE XC functionals.
Standard Troullier-Martins pseudopotentials~\cite{Troullier-Martins} 
from the ABINIT library were used for both.  
Geometries were optimized using 
the Broyden-Fletcher-Goldfarb-Shanno algorithm,~\cite{Schlegel} 
to a force tolerance of $5\times10^{-5}$~hartree/bohr.  

The main convergence error in our calculations was that of using
a finite-sized periodic simulation cell, necessitated by the use
of a plane-wave expansion.  The simulation cell size was chosen so
that total energies were converged to within $3\times10^{-5}$~hartree.
Errors in nearest-neighbor bond-lengths due to finite system size 
are less than $5\times10^{-5}$~\AA.
In order to get good spatial resolution of plots, 
we took a plane-wave cutoff of 99 hartree for all systems, 
well above that needed for convergence
of energies to chemical accuracy ($<\! 40$~hartree) in the 
pseudopotential systems. 
The convergence errors in total energy from the 
finite plane-wave cutoff range from 10$^{-7}$ hartree for $\SiHfour$ to 
10$^{-6}$ for $\CfourHeight$ and the error in nearest-neighbor bond-lengths, 
from 10$^{-7}$ to 10$^{-6}$~\AA. 
Converged simulation-cell parameters for each molecule may be
found in the supplementary information.

Given a periodic cell, the density and related expectations should suffer 
boundary effects.  Most notably, whereas the density and its derivatives and 
the kinetic energy density should decay exponentially to zero in 
a finite system, these will approach a small finite value at the cell 
boundary.
For the cell sizes 
used, this minimum value of the density is on the order of 10$^{-8}$ a.u., 
for systems with maximum densities on the order of an a.u.;
a significant distortion from exponential decay is observable only within 
two bohr of the location of the minimum.

The ABINIT code outputs density and KE density 
as a three-dimensional uniform grid over the periodic simulation cell, with 
grid spacing determined by the dimensions of the fast Fourier transform %(FFT) 
used in the plane-wave code.  The real-space grid %inverse to the FFT grid 
used to accommodate a 99 hartree plane-wave cutoff has a resolution of 
0.11 bohr, defining the resolution of our plots.
The Laplacian and gradient of density were
evaluated numerically on this grid using a 
Lagrange-interpolating polynomial method.
Color surface plots and contours were generated using gnuplot and the
associated pm3d utility. 

\section{Results \label{sec:results}}

First of all, to assess the quality of data within the plane-wave
pseudopotential approach, we show 
results for basic structural properties for the AE6 test set. 
Table~\ref{table:bond length} shows the 
mean relative error (MRE) and mean absolute relative error (MARE) 
of LDA and PBE pseudopotential predictions of bond lengths for the AE6
test set, as compared to experimental data.
The LDA gives an excellent fit to double and triple bonds and about a 
1\% over-binding of single bonds, in line with other results for the
LDA.~\cite{AE6,Staroverov,SOGGA} An atypical tendency to under-bind for C--H 
bonds leads to a MRE whose accuracy we suspect would not hold for 
larger test sets.
The overall tendency of the PBE is to increase bond lengths
relative to the LDA, again the expected trend, which results in a slightly 
better absolute agreement with experiment.

\begin{table}[ht]
\begin{tabular}{ccc}
\hline
\hline                 & LDA    & PBE \\
\hline \hline MRE (\%)  & -0.006 & -0.087 \\
       \hline MARE (\%) &  0.68  &  0.53  \\
\hline \hline
\end{tabular}

\caption{
Performance of pseudopotential DFT calculations for the
bond lengths of the AE6 test set -- 
mean relative error (MRE) and absolute relative error (MARE) in angstroms
compared to experimental data from Ref.~\onlinecite{CCCBDB}.
\label{table:bond length}
}
\end{table}

The summary performance of DFT predictions for atomization energies 
is shown in Table~\ref{table:atomizationenergy}. 
Again the 
trend of the LDA is to over-bind with respect to experiment and that
of the PBE to %relax binding and 
remove much of this error. 
The LDA does worst energetically for systems with a double or 
triple bond: S$_2$, SiO and C$_2$H$_2$O$_2$.  
Our pseudopotential estimate of the MAE for the PBE functional 
on the AE6 test set compares reasonably well with those obtained from
all-electron calculations~\cite{HTBS,RGE2} using
gaussian basis sets.
The two approaches agree for singly-bonded systems
while our pseudopotential approximation overestimates the 
atomization energy of molecules
with double bonds by about 10~kcal/mol per double-bond.  
A purely numerical calculation on an ultrafine grid~\cite{SOGGA}
%basis-set free 
reports a MAE of 3.0~kcal/mol per bond for the PBE functional as compared 
to 3.6~kcal/mol per bond here, indicating that use of a pseudopotential 
overestimates binding but perhaps not by as much as indicated by the 
gaussian basis-set calculations.
In any case, this error is 
minute in comparison to the large errors between orbital-free and 
Kohn-Sham kinetic energies.

Further information about the convergence with respect to the finite size
of the cell is shown in the supplementary material for this paper,~\cite{supplementalfootnote}
including converged finite-size cell parameters for each molecule 
in Table~S-I and finite cell boundary errors for $\Stwo$ in Fig.~S-1.
Table~S-II shows per-molecule data from LDA and PBE pseudopotential 
calculations of the bond lengths of the AE6 test set,
compared to experimental data, and S-III does the same for atomization 
energies.

\begin{table}[t]
\begin{tabular}{cccc}
\hline\hline
                 & LDA      & PBE   & PBE-ae \\
\hline
MSE               & 67.1 &   20.8 & 12.0 \\
MAE               & 67.1 &   23.0 & 15.1 \\
MARE (\%)         & 16.0 &    7.4 & 4.4 \\
\hline\hline
\end{tabular}
\caption{\label{table:atomizationenergy} 
Summary errors (mean signed, mean absolute and mean absolute 
error in percent) of pseudopotential DFT calculations, and of 
an all-electron PBE calculation~\cite{HTBS} for the
atomization energy of the AE6 test set, measured relative to 
experimental data from Ref.~\onlinecite{AE6}. In kcal/mol.
}
\end{table}

\subsection{Electronic structure: atoms \label{sec:psptest}}
\begin{figure}
\includegraphics[width=0.40\textwidth]{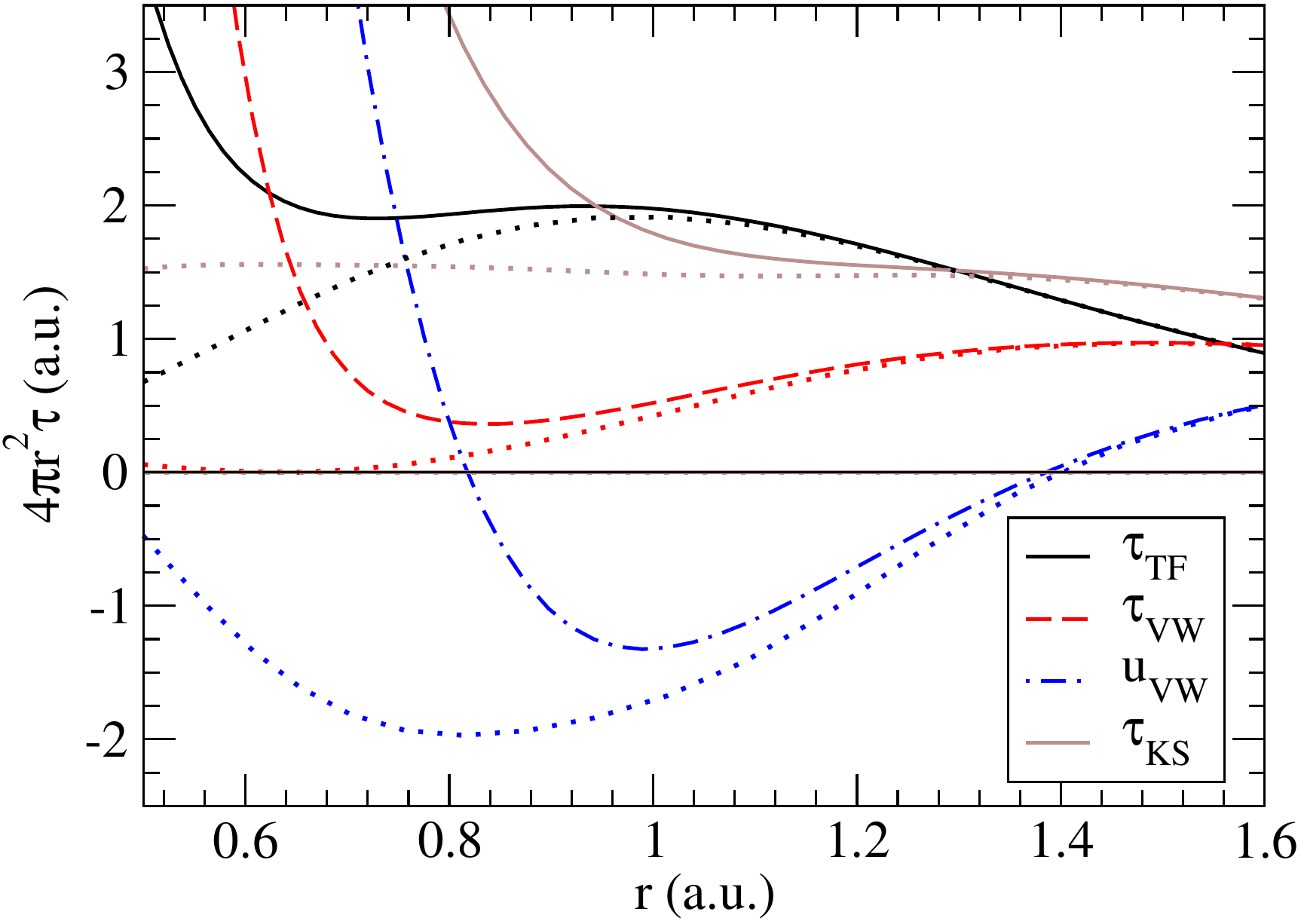} %.pdf}
\caption{\label{fig:psptest}
(color online)
Comparison of all-electron and pseudopotential kinetic energy
densities for the carbon atom.  
Shown are the radial probability density versus distance
from nucleus for the Thomas Fermi (solid line), von~Weizs\"acker (dashed), 
and Kohn-Sham KED (lighter solid) as well as the quantity $\uvW$ (dot-dashed) 
defined in the text.  The equivalent pseudopotential quantity for each is shown
as a dotted line, matching at the cutoff radius 1.498~$a_B$.
}
\end{figure}

Before showing results for molecules, it is instructive to compare
pseudopotential and all-electron results for atoms.  Fig.~\ref{fig:psptest}
demonstrates this comparison for the density, its gradient and Laplacian and 
the Kohn-Sham KED of the C atom.
In order to make a clean comparison between quantities, we 
convert the first three functions into equivalent kinetic-energy density 
models:  $\tauTF\!\sim\!n^{5/3}$, $\tauvW \equal \gradnsq/8n$, and 
$\uvW \equal \lapln/4$.
The last is generated by taking the functional derivative of 
$\tauvW$ with respect to density.

The pseudopotentials for C are designed so that 
the pseudo-valence density matches the 
all-electron density after a cutoff radius of 1.498~$a_B$.  This match
is respected for the other quantities as well.  However the
pseudodensity and thus $\tauTF$ continue to match the real density 
with reasonable agreement almost to the core-valence transition radius at 
about 0.8~$a_B$.  
%Inside this radius, the pseudo-valence charge density 
%smoothly dips to a minimum at the nucleus.  
The other pseudo-quantities deviate from their all-electron equivalents much 
more quickly, especially $\tauKS$ and $\uvW$.  
In the all-electron case, core orbitals smooth out the density and thus
reduce the magnitude of the $\lapln\!<\!0$ peak in the valence shell, 
and they add extra terms to $\tauKS$.  
The quantitative impact on $\lapln$ is quite significant:
the region of peak negative $\lapln$ (the valence shell charge concentration
or VSCC in QTAIM analysis) is 
broader in extent and the position of the critical point about 40\% closer 
to the nucleus than in all-electron calculations.  The maximum negative 
value of $\lapln$ is typically three times larger than its all-electron 
equivalent, with similar errors for the VSCC's of molecules.  
Plots shown below for $\lapln$ and $\tauKS$ in molecules do %, while they
faithfully reproduce the qualitative topological features of the 
all-electron case, and are quantitatively accurate at bond centers and
asymptotically; but they must be treated with caution with respect to 
other quantitative details. 

\subsection{Electronic Structure: molecules \label{sec:aesix}}
\begin{figure*}[tp]
\centering
\includegraphics[]{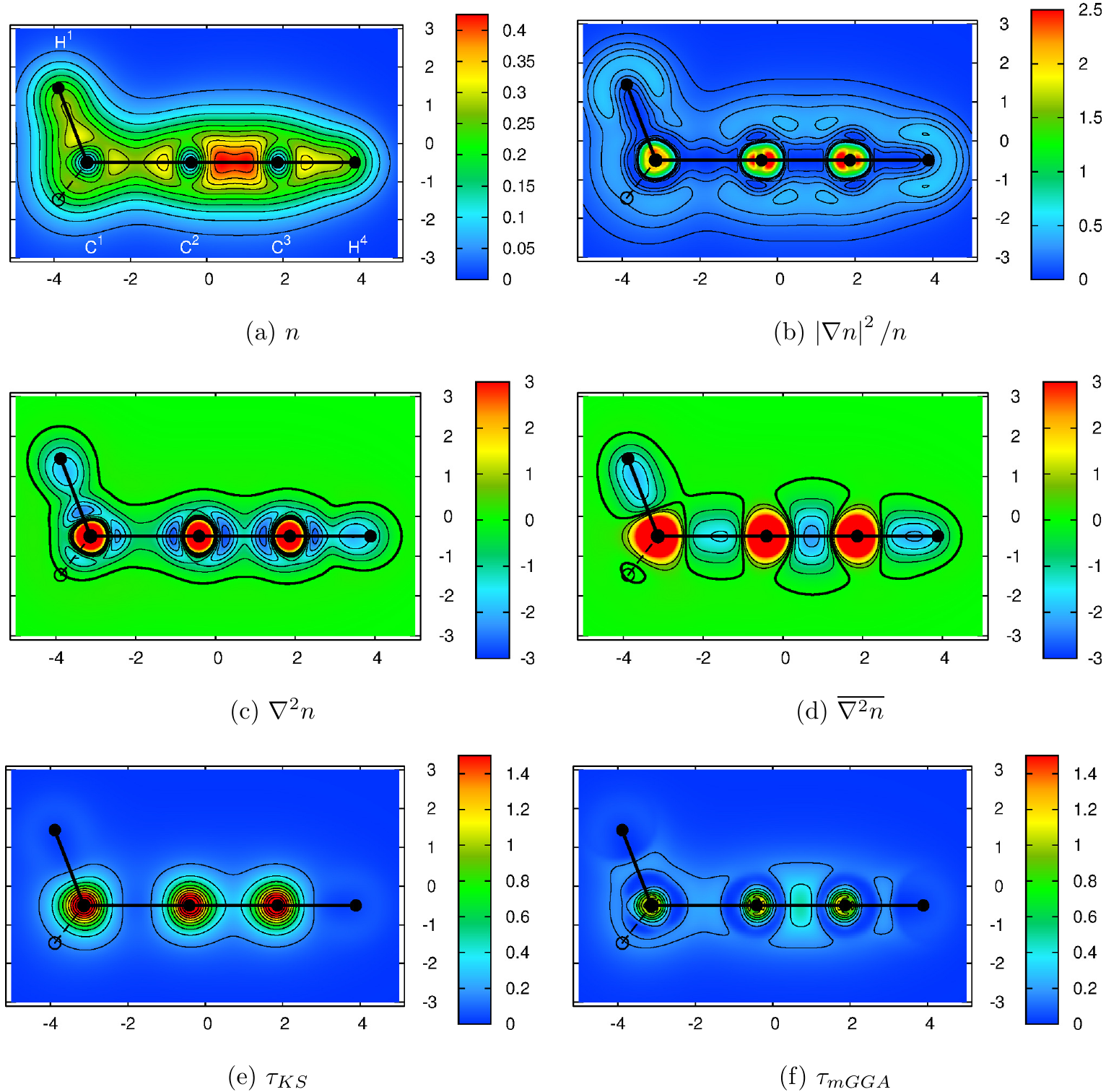} %.pdf}
\caption{
Functionals of the density for $\CthreeHfour$ within the pseudopotential
approximation, showing cut through the C--C--C bond axis and two hydrogens. 
In-plane atoms and bonds are shown as black disks 
and line segments; specific atoms are identified by labels.  
Out-of-plane ones shown as dashed lines and open disks.  Contour
levels for Laplacian (c)  and pseudo-Laplacian of Eq.~(\ref{eq:laplgea}) (d) 
are identical, with thick contour at zero.  Contour levels for mGGA KED (f)
are the same as those of the Kohn-Sham KED (e). 
}
\label{fig:C3H4ppr}
\end{figure*}

\begin{figure*}[tp]
\centering
\includegraphics[]{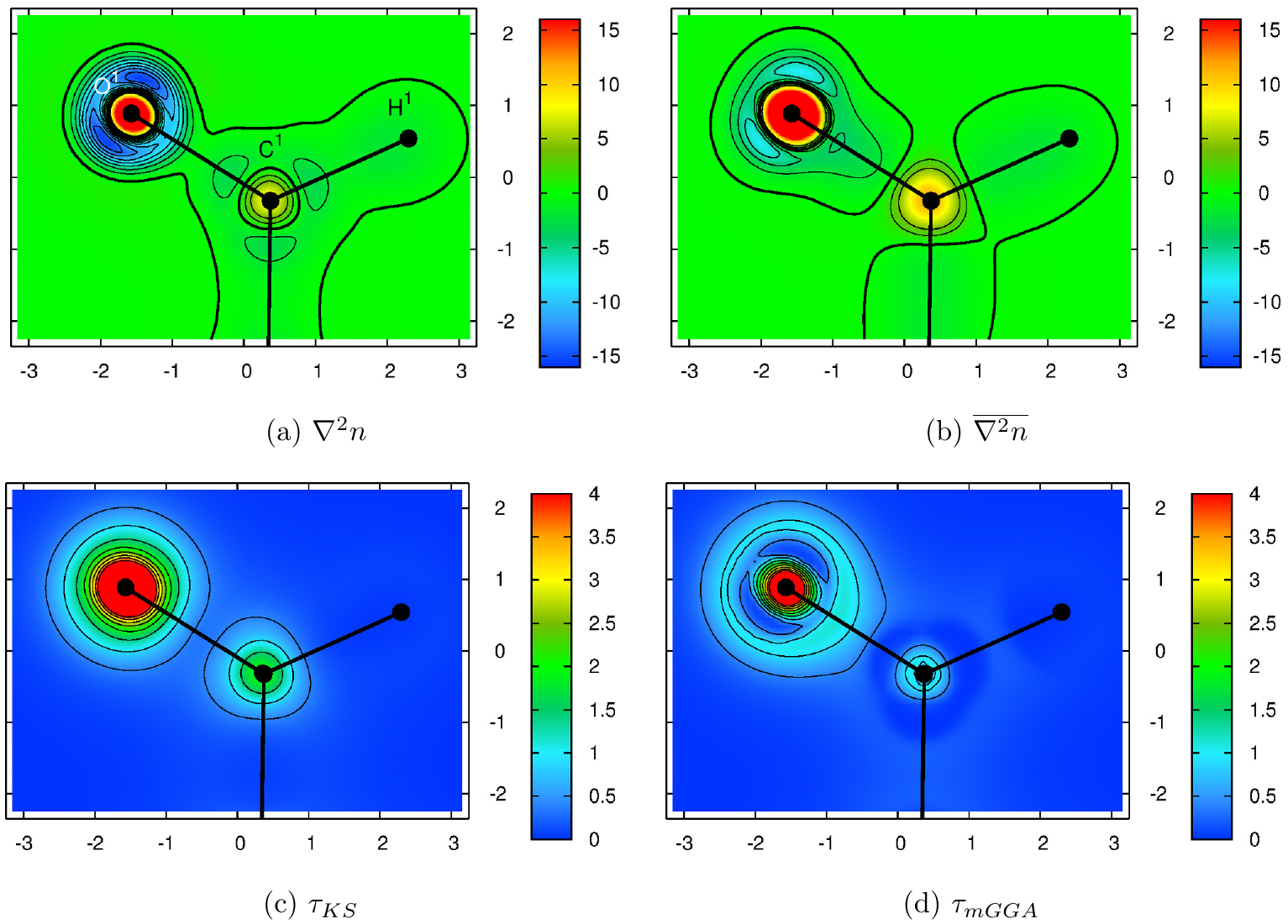} %.pdf}
\caption{
Functionals of the density for pseudo-$\CtwoHtwoOtwo$, showing cut through the bond
plane and an oxygen, carbon, and hydrogen atom.  
Details are the same as in Fig.~\ref{fig:C3H4ppr}.
}
\label{fig:C2H2O2ppr}
\end{figure*}

\begin{figure*}[tp]
\centering
\includegraphics[width=0.75\linewidth]{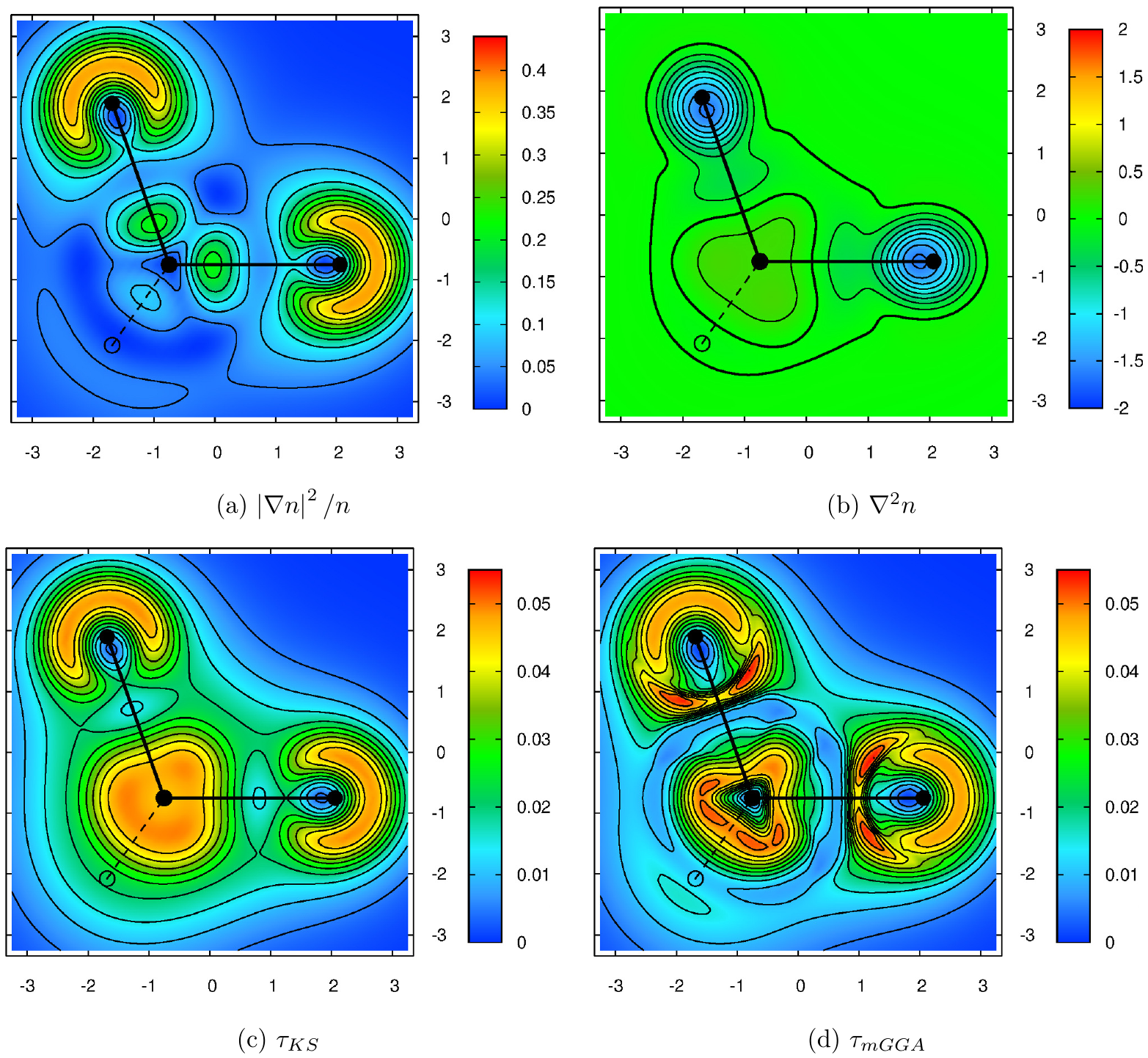} %.pdf}
\caption{
Functionals of the density for pseudo-$\SiHfour$, showing cut through 
a plane containing a Si atom at center and two tetrahedrally bonded hydrogens.
Details are the same as in Fig.~\ref{fig:C3H4ppr}.
}
\label{fig:SiH4ppr}
\end{figure*}

\begin{figure*}[tp]
\includegraphics[width=0.42\linewidth]{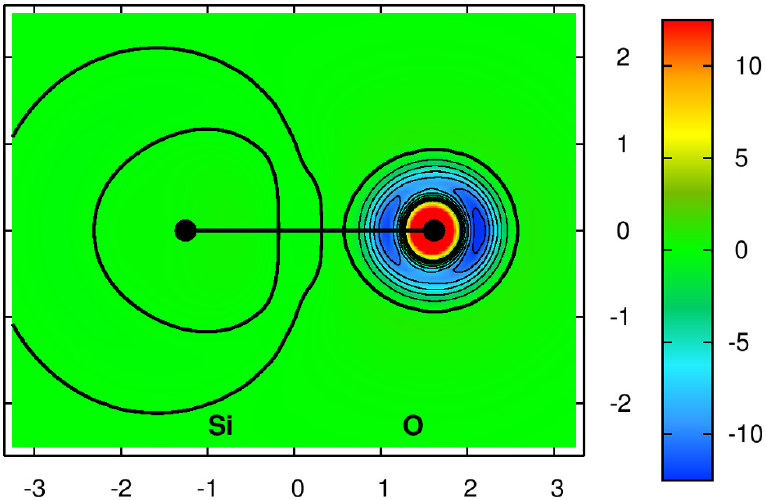} %.pdf}
\caption{
Laplacian of the density for pseudo-SiO, showing cut through the bond 
axis.  Details same as for Fig.~\ref{fig:C3H4ppr}(c). 
}
\label{fig:SiOsht}
\end{figure*}

Figures~\ref{fig:C3H4ppr},~\ref{fig:C2H2O2ppr}, \ref{fig:SiH4ppr} and 
\ref{fig:SiOsht} show contour
plots for the density and related quantities for pseudopotential models
of several of the 
molecules of the test set: $\CthreeHfour$, $\CtwoHtwoOtwo$ and $\SiHfour$.
and SiO.  
In Fig.~\ref{fig:C3H4ppr}, we show in the first row 
(a) the ground-state pseudo-density $n$ and 
(b) the gradient factor $\gradnsq/n$ 
that appears in the gradient expansion of $\tauKS$ [Eq.~(\ref{eq:taugea})] 
and the von-Weizs\"acker KED [Eq.~(\ref{eq:tauvw})].
The second row shows
(c) the Laplacian of the density $\lapln$, and the gradient-expansion derived
pseudoLaplacian [Eq.~(\ref{eq:laplgea})] used in metaGGA's.  The third row 
shows (e) the Kohn-Sham KED $\tauKS$ and (f), the Perdew-Constantin mGGA
model for the same.  All quantities are plotted in hartree atomic units;
all except (a) are thus dimensionally energy densities.
The other figures show subsets of this suite of data, as identified by
subcaptions, for the other three molecules.

Each figure shows a 
two-dimensional slice through the molecule, with a color surface
plot with values ranging from blue (minimum value shown) to red (maximum).
The numerical scales for the surface plots are shown in the bar to the right
of each subplot.  Superimposed upon these are contour plots. Thicker 
contour lines for the Laplacian and 
pseudo-Laplacian [Fig.~\ref{fig:C3H4ppr}(c) and (d)] indicate the zero contour; the other four functions plotted are positive definite.  The contours are 
adjusted to bring out details of bonding regions, and do not cover the atom 
cores.
Contour values and ranges for the equivalent quantities of Laplacian
and pseudo-Laplacian are identical, as are those for 
the two KED's in subplots (e) and (f). 
Atoms and bonds in the plane of a plot are indicated by black dots and thick
black lines; projections of out-of-plane atoms and bonds onto the plane of the
plot are shown as open circles and thick dashed lines.

We start out with a discussion of the four direct measures of
electronic structure -- the density, $\gradnsqn$, $\lapln$, and $\tau$ and 
consider the two approximations $\barlapln$ and $\taumGGA$ 
in following subsections. 

The subject of Fig.~\ref{fig:C3H4ppr}, 
$\CthreeHfour$,  
is perhaps the richest example of the test set,
illustrating several types of bonds.  Three hydrogens (H$^1$ to H$^3$) bond 
with a tetrahedral geometry to a carbon (C$^1$), which is joined to the second 
carbon ($C^2$) through a single bond, the second carbon shares a 
triple bond with the third (C$^3$), and this is terminated with 
a final carbon-hydrogen bond.
Our plots show a cut through the three central carbon pseudo-atoms aligned 
along the $x$-axis and one hydrogen on either end; the other two hydrogens 
extend out on either side of the far left-hand side of the plane. 

The valence particle density (a) shows some features common to each
molecule of the set.  
As in the single-atom case (Fig.~\ref{fig:psptest}), the density tends 
smoothly to a minimum at the center of each carbon 
pseudo-atom.  In contrast, the hydrogen atom has no core electrons and the 
effect of the pseudopotential is simply to smooth out the cusp in the 
density at the nucleus.
The highest electron density is thus naturally within bonds -- 
especially the triple bond ($\mathrm{C}^2-\mathrm{C}^3$). 
The gradient-squared of the density (b) is nearly zero 
in regions with bonding, where the Laplacian (c) shows most structure, and is
largest in the pseudo-atom core and at the edges of the molecule where 
$\lapln$ is zero, as indicated by the thicker contour line in (c).  The 
Laplacian is negative almost entirely along the center 
except for the interior of each carbon pseudo-atom.  
This is a hallmark of covalent bonding in 
QTAIM analysis~\cite{BaderReview,BaderEssen84}  -- 
the center of a bond is a saddle critical point for the particle 
density, with a negative value for a covalent bond because of the 
the buildup of charge between atoms. 
The value of $\lapln$ at the C$^1$--C$^2$ bond critical point is -0.700 
and that for C$^2$--C$^3$ is -1.143, reasonable values for C--C bonds.~\cite{BaderEssen84}
$\lapln$ is positive in the pseudo-atom core, where the density is 
at a local minimum, and in the classically forbidden region far from 
the molecule.
The Kohn-Sham kinetic energy density (e), 
is the smoothest and least structured of the
measures of the Kohn-Sham system shown. %%in the figure.
As its relationship to the electron density is nontrivial,
it not surprisingly appears to have little apparent correlation with it.
It is primarily concentrated in the pseudo-atom cores with a strong
peak at the center of the pseudo-atom.  This follows the qualitative trend
of the KED of all-electron systems,~\cite{JacobsenPCCP2013} except for the 
absence of shell structure.  
Otherwise it is significantly larger in the triple bond than 
in the single bonds, where it is nearly zero. 

Fig.~\ref{fig:C2H2O2ppr} shows Laplacian and KED quantities
for the $\CtwoHtwoOtwo$ pseudo-molecule.
This has a trigonal-planar form with a line of symmetry
through the center of the molecule.  
The oxygens share polar double bonds with 
the carbons, 
and the 
carbons form covalent single sp$^2$ bonds with each other and the hydrogens. 
There are two lone pairs of electrons present on each oxygen. 
The plot shows one oxygen, carbon and hydrogen, and part of the C--C bond 
at the bottom of the plot.
The single C--C and C--H bonds are very similar to those in $\CthreeHfour$,
so that the scale is adjusted to favor the oxygen atom which has a much
larger density and KE density.  

The C--O bond, being polar, exhibits several 
features not seen in $\CthreeHfour$.
The density gradient is nonzero in the bond -- the push of 
density towards the more electronegative oxygen causes a local saddle point 
in the gradient on the oxygen side.  
VSCC lobes due to two sets of unpaired electrons are 
identifiable on the oxygen, but none on the bond axis,
a reflection of the change in character of the bond. %seen in the gradient.
%The minimum value of $\lapln$ at the bond center is close to 
%zero as indicated by a notch in the zero contour.  
However, the VSCC lobes of peak negative $\lapln$ (a) 
around the carbon atom are similar to those of the pure covalent bond.  
The kinetic energy density, as for $\CthreeHfour$, is concentrated in 
atom cores with little contribution from within the bond, and thus
the bond's polar character causes no observable change from that of the covalently bonded system.

Next we consider $\SiHfour$, 
a nearly spherical molecule closely resembling a filled-shell atom 
in structure.  
It exhibits straightforward tetrahedral
bonding, with sp$^3$ hybridization of the silicon orbitals and 
covalent Si-H bonds.  
Fig.~\ref{fig:SiH4ppr} shows a cut through a plane
containing three of the atoms (H, Si, H) of the pseudo-molecule.  
A pair of hydrogen atoms is located above and below the plot plane as
indicated by the open circles.

An item of interest is the comparison between $\tauKS$ (c) 
and $\gradnsq/n$ (a).  Recalling that the von~Weizs\"acker KED 
[Eq.~(\ref{eq:tauvw})] is $\gradnsq/8n$,
we set the color scales of (a) and (c) to an exact 8:1 ratio
so that a comparison of $\tauKS$ relative to $\tauvW$ can be made.
(For the other molecules, such a scheme wipes out almost all 
information about the gradient of the density, because $\tauKS$ is 
much larger than $\tauvW$.)  Here it is apparent that the Kohn-Sham KED reaches
the von~Weizs\"acker limit everywhere in the vicinity of a  
hydrogen atom.  This seems reasonable in that each hydrogen atom
has a single occupied orbital, and is in a sense a paradigm for 
the von~Weizs\"acker limit in fermionic systems.
The Laplacian for the Si--H bond (b) 
heavily emphasizes the H atom 
because of the more dispersed nature of Si valence orbitals as compared to those
of H.

A final example from the test set is SiO, which features a double bond
that should be polar covalent given an electronegativity difference of 1.6. 
Fig.~\ref{fig:SiOsht} shows a surface plot 
of the Laplacian of the density %and the pseudo-Laplacian (b) 
for a cut through the pseudo-molecule Si--O bond.
Other quantities are available for SiO in the supplementary material.
The valence electrons that participate in this bond heavily favor 
oxygen, the more electronegative atom,
leaving the silicon atom hypovalent.~\cite{CCCBDB}  
%The negative lobes in the Laplacian are aligned on axis, indicating sp rather than sp$^2$
%hybridization in the formation of molecular orbitals, with one lone 
%pair on the oxygen.
%Two $\pi$ bonds make the molecule axially symmetric.
%
The zero contour (thicker contour) of $\lapln$ is of interest
for this system as it indicates a bond topology qualitatively different from 
the other cases.  The orientation of the zero contour, crossing 
the bond perpendicularly to the bond axis,
indicates that $\lapln\!>\!0$ in the bond center -- specifically
in the region between the two closed contours surrounding Si and O respectively.
This indicates in QTAIM analysis that the bond is ionic; the maximum
value of $\lapln\!=\!0.194$ is comparable to that of NaCl.~\cite{BaderEssen84}
There is essentially a catastrophe in the topology of the zero 
contours whereby the Si--O bond cannot be mapped to a 
polar-covalent geometry, such as the C--O bond in $\CtwoHtwoOtwo$, without 
breaking and rejoining contours.  
%Such a topological change in $\lapln$ is also a characteristic of the 
%dissociation
%of covalent bonds into dissociated reagents,~\cite{BaderReview} and can
%be seen~\cite{Tachibana2001} for the KED defined with the 
%Schr\"odinger convention [Eq.~(\ref{eq:tauksp})]
%which involves the Laplacian of Kohn-Sham orbitals.

Surface plots for the last two molecules of the AE6 test set 
($\Stwo$ and $\CfourHeight$) are not shown in this 
paper but are available in the supplementary data.~\cite{supplementalfootnote}  
Although they repeat themes already discussed for the other molecules, 
they have individual characteristics that 
should benefit from further investigation.
The triplet
ground state of $\Stwo$, with a double bond and two lone pairs per atom, 
is similar in structure to $\CthreeHfour$ and $\CtwoHtwoOtwo$.
Nevertheless, 
%it is the one system in the test set where 
the KED shows interesting structure  
near the valence shell peak and in the bond.
Cyclobutane ($\CfourHeight$) is a cyclic molecule with a ring of four carbons
and two hydrogens bonded to each. %carbon as well.
Unique to this system is the low-density region inside the carbon ring 
where the gradient of the density is zero but the Laplacian and KED are not.  
This topology is similar to that of the bond-center of a nearly dissociated 
molecule, and not found elsewhere (at equilibrium geometry)
in the test set. %evaluated at equilibrium geometry. 
Such regions have been of interest for QTAIM analysis~\cite{BaderReview} 
and may provide a glimpse into how well approximated KED's 
perform in predicting binding.

\subsection{Gradient expansion for the Laplacian\label{sec:barlapln}}

The subfigure (d) of Fig.~\ref{fig:C3H4ppr} 
and (b) of Fig.~\ref{fig:C2H2O2ppr}    %and~\ref{fig:SiOsht} 
show the pseudo-Laplacian $\barlapln$ [Eq.~(\ref{eq:laplgea})] which 
approximates the Laplacian in 
terms of the electron density, its gradient and the kinetic energy density.
Up to a small correction proportional to $\gradnsq/n$, this quantity
is simply $6(\tauKS \!-\! \tauTF)$; given that $\tauTF$ is 
a power of the particle density, it interprets
the Laplacian as roughly a measure of the difference
between the kinetic energy and particle densities.
As seen especially in Fig.~\ref{fig:C3H4ppr}, our data support 
this qualitative picture.
The Laplacian (c) is positive and large in the carbon pseudo-cores, precisely
where the kinetic energy density (e)  is largest and the charge 
density (a) is at a minimum; it is most negative in the bond regions where 
the situation is reversed.  As a result, 
$\barlapln$, plotted in (d) with the identical set of contours as $\lapln$,
captures the basic qualitative trends of this quantity, and on average,
its relative magnitude in each bond.  
In contrast, the zero contour of $\lapln$ and $\barlapln$,
shown as thicker black contours, have qualitatively different topologies.
However, it seems reasonable to expect that, 
given their qualitative similarity, they could produce similar results
if used as parameters in a functional for 
an integrated quantity such as the exchange energy. 
Notably, the contour of $\barlapln\!=\!0$ closely matches the shape of the 
$1/2$-contour of the LOL, a close equivalent when density gradients
are small.~\cite{JacobsenPCCP2013}

%The pseudo-Laplacian is also shown for several other cases, where
%similar trends may be seen.  The model is almost exact for $\Stwo$,
%with nearly perfect quantitative prediction of $\lapln$ in the valence
%region and a nearly identical zero contour. 
%It is at its worst for the molecules containing oxygen, where it severely 
%under predicts the magnitude of the Laplacian in the oxygen valence shell. 
%The marginally ionic SiO shows the worst agreement between pseudo-Laplacian 
%and Laplacian.  
%The ionic bond comes out as
%covalent if we apply the $\lapln$-based criterion of 
%QTAIM~\cite{BaderEssen84} using $\barlapln$.

A check on the quality of this approximation can be obtained
by the sum rule for $\lapln$.  Since it is an exact derivative,
the integral of $\lapln$ over the entire unit cell should be exactly zero.
While the integral for $\lapln$ is zero to within round-off error, 
that for $\barlapln$ ranges from about
0.1 hartree for $\SiHfour$ to about 20 hartree for the largest molecules.
This is a reflection of the very large difference between the 
integrated Kohn-Sham kinetic energy and that of the Thomas-Fermi approximation.
Energy densities can differ by several orders of magnitude in the 
pseudo-atom cores, an effect beyond the scale of our surface plots, 
but clearly shown in the log plots in Sec.~\ref{sec:lineplots}.  
%Thus, the pseudo-Laplacian 
%will not serve as a quantitative replacement for the Laplacian in a metaGGA
%without a substantially different treatment of the large-inhomogeneity
%limit.

\subsection{The mGGA model for the kinetic energy density\label{sec:KED}}

The final quantity for which we have made surface plots 
is the mGGA orbital-free KED.~\cite{PerdewConstantin}
%of Perdew and Constantin,~\cite{PerdewConstantin} 
%a metaGGA level 
%orbital-free model for the Kohn-Sham kinetic energy density.  
It is shown for three molecules, subfigure (f) of 
Fig.~\ref{fig:C3H4ppr} and (d) of Fig.~\ref{fig:C2H2O2ppr} and~\ref{fig:SiH4ppr},  
with contours and color scale that duplicate those of 
the Kohn-Sham KED. 
The agreement between the two is generally not good.  For $\CthreeHfour$ 
(Fig.~\ref{fig:C3H4ppr}) 
the mGGA, like the true KED, peaks in the pseudo-atom core, but is much  
smaller in magnitude.
It is too large in the
center of bonds, particularly the C$^2$--C$^3$ multiple bond. 
%s like that between C$^2$ and C$^3$.
The most striking difference
%between the two 
is the dramatic drop in magnitude 
in the mGGA in the region of peak valence charge concentration surrounding 
each carbon atom.  The shape of these zeroed-out regions 
correlates with the VSCC lobes of the Laplacian %(of peak negative $\lapln$) 
accentuating regions of peak density.  The identical pattern shows up 
in $\CtwoHtwoOtwo$ 
(Fig.~\ref{fig:C2H2O2ppr}), with the KED zeroing out in VSCC regions
for both oxygen and carbon,
almost perfectly matching
the contours of $\lapln$ for the two lone oxygen pairs. %of the oxygen atom.
This pattern occurs around the carbon atoms of $\CfourHeight$, 
the two lone pairs of each S atom in $\Stwo$ and of the oxygen atom of SiO,
indicating a global trend. 

$\SiHfour$, shown in Fig.~\ref{fig:SiH4ppr}, is a case in which 
the mGGA works.  In this case, much of the
system is already very close to the von~Weizs\"acker limit, which the 
mGGA is designed to capture exactly.  
Moreover, errors in the mGGA in different regions,
such as in the Si atom core and near the hydrogen atom, almost exactly
cancel,
leading to a qualitatitively much better match of the mGGA to the exact
KED than for the other five cases.
(Notably, the Si pseudo-atom lacks the strong VSCC lobes associated
with unusually low mGGA KED in $\CthreeHfour$ and $\CtwoHtwoOtwo$.)
%(The regions of excess KED near the hydrogen atom
%in the mGGA are correlated to a notch in the zero contour of 
%$\lapln$, indicating a region of
%smaller than normal charge concentration -- the reverse of the 
%condition for excessively small KED in this model.)

\subsection{Plots through bond axes\label{sec:lineplots}}
In this section, we focus on the quantitative comparison of  
various models for the kinetic energy density, for which 
linear plots are convenient. 
We plot the enhancement 
factor $F_S = \tau/\tauTF$, which avoids excessive differences in scale 
between atoms.
%as each KED can have large differences in scale
%from atom to atom, but their ratio $F_S$, much less so. 
We are also interested in the measure of electron localization $\alpha$ 
[Eq.~(\ref{eq:alpha})], 
that can also be thought of as
the Pauli contribution to the enhancement factor. 
%-- the difference between it and the von-Weizs\"acker $F_S$. 

\begin{figure}
\includegraphics[width=0.42\textwidth]{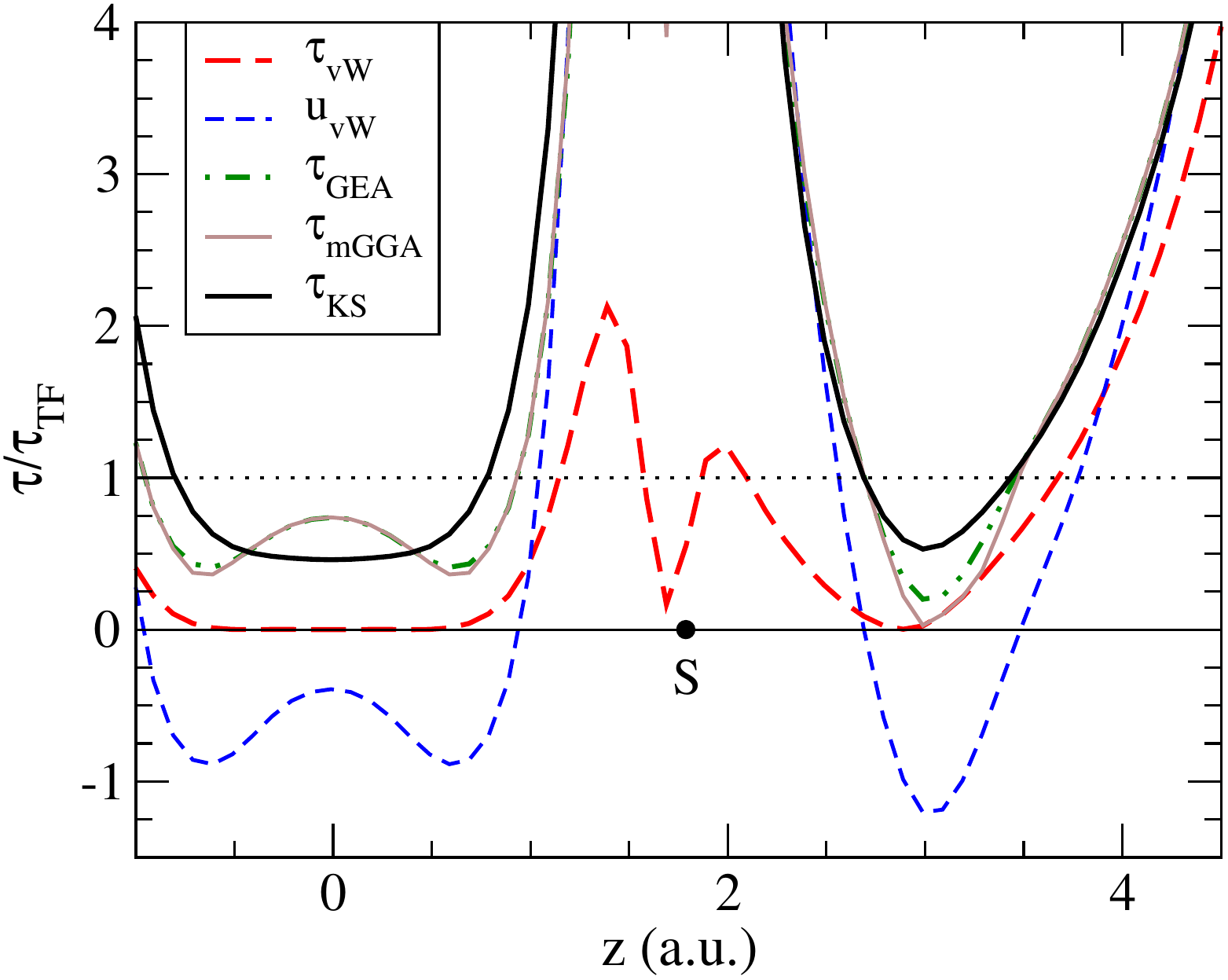} %.pdf}
\caption{\label{fig:S2fkden}
(color online)
The enhancement factor $F_S = \tau/\tauTF$ for 
the Kohn-Sham kinetic energy density (solid black line) and 
various orbital-free models, within the pseudopotential approximation,
versus position along the bond axis $z$ for $\Stwo$.  %the disulfur molecule.  
Also shown is $F_S$ for $\uvW\!=\!\nabla^2 n/4$. 
The Thomas-Fermi result, trivially one, is shown as a dotted line.
Location of the sulfur atom noted by solid dot.
}
\end{figure}

In Fig.~\ref{fig:S2fkden} we show $F_S$
of several model KED's for the pseudopotential approximation to the 
disulfur molecule $\Stwo$ 
as a function of displacement $z$ from the molecule center along the bond axis. 
The focus is on a single sulfur pseudo-atom, marked by the black dot on the 
$F_S\equal 0$ axis;
the molecule has a mirror-symmetry plane through the bond center at $z\!=\!0$.
The Thomas-Fermi result is  %, by the definition of $F_S$, 
the horizontal line $F_S\!=\!1$.  
The von Weizs\"acker enhancement factor, $\tauvW/\tauTF\!=\!5p/3$,
%is proportional to $\gradnsqn^{8/3}$; it 
is  nearly zero in the bond region and again at the density peak associated
with the lone pair behind the bond.  The related expectation 
$\uvW$  %%%% \!=\! \nabla^2 n/4$ %(short dashed line)
has an enhancement factor equal to $10q/3$. 
It is negative 
in the covalent bond and the lone-pair behind the sulfur atom, and positive
in the pseudopotential core and asymptotically.

%$\tauKS$ is positive definite by construction, and although it 
Both the gradient expansion $\tauGEA$ and the more sophisticated 
$\taumGGA$ are positive definite, in agreement with the required behavior
of $\tauKS$.
%are positive definite for $\Stwo$, as required for $\tauKS$.   
%Each of the three is at 
Each is at
a minimum in the bond and lone-pair regions, reach local maxima in the
core and tend to $\infty$ asymptotically. 
However, $\tauKS$ is smooth and featureless, lacking the 
oscillatory structure of the gradient and Laplacian of the density.
The mGGA, where it differs from the GEA, does a slightly worse
job in describing the Kohn-Sham value.  In the lone-pair region
around $z\!=\!3$~a.u., it suffers from the extinction effect seen in the 
surface plots for $\CthreeHfour$ and $\CtwoHtwoOtwo$.  
Here, $u_{vW}/\tauTF \!<\! -1$, equivalent to $q \!<\! -0.3$, 
which proves to be a significant criterion for this problem to occur 
in the mGGA.
Overall, the mGGA fares better for $\Stwo$ than for other molecules, 
perhaps because this error in its enhancement factor is cancelled
by a reverse effect at the center of the double bond.
The electron-localization measure $\alpha$, not shown in 
Fig.~\ref{fig:S2fkden}, is available in Fig.~S-1 of the 
supplementary material.~\cite{supplementalfootnote}
%The value of $\alpha$ in the double bond is 0.5, typical of multiple bonds, 
%and slightly smaller behind the lone pair.

\begin{figure}
\includegraphics[width=0.43\textwidth]{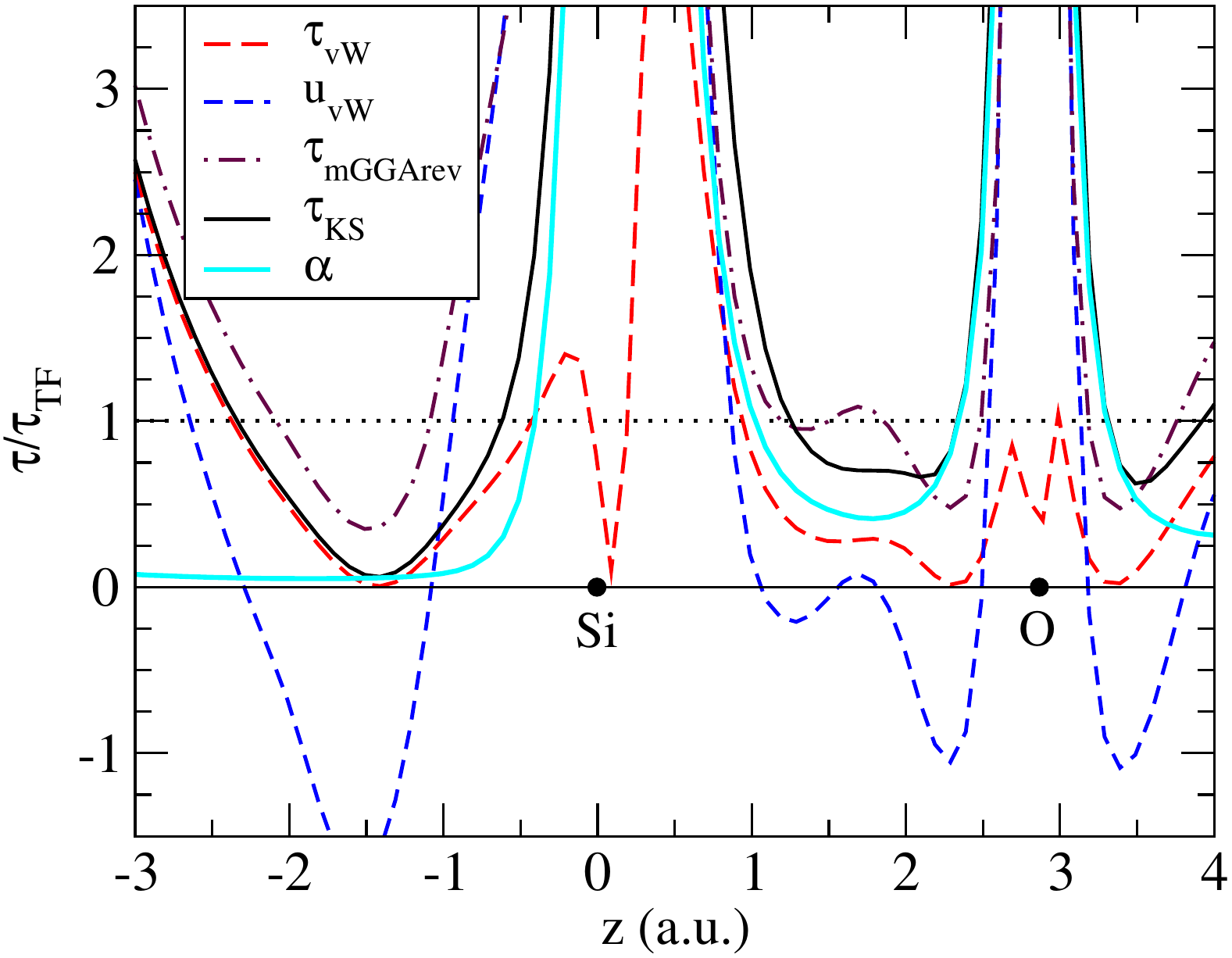} %.pdf}
\caption{\label{fig:siofkden}
(color online)
The enhancement factor $F_S\! =\! \tau/\tauTF$ for 
the Kohn-Sham and von Weizs\"acker KED's, compared
to the difference between the two, $\alpha$,
versus position along the bond axis $z$ for pseudo-SiO.
Also shown is $F_S$ for $u_{vW} \sim \lapln$ and the mGGArev discussed
in Sec.~\ref{sec:mggarev}; dotted line %at $F_S\!=\!1$ 
indicates the Thomas-Fermi result.
Location of each atom on bond axis noted by a solid dot.
}
\end{figure}

Fig.~\ref{fig:siofkden} shows enhancement factors for
pseudo-SiO.  As noted previously, this is the most polar molecule
in the test set and gives a structural contrast
to the more covalent molecules.  As such we focus on $\tauvW$ and $\uvW$ 
as stand-ins for the gradient and Laplacian of the density,
and related variables $p$ and $q$, as compared to 
the Kohn-Sham KED.
The gradient squared of 
the density ($\sim\! \tauvW$) does not vanish in the bond, as the 
density steadily increases from 
the Si valence shell to the O.  The Laplacian ($\sim\! u_{vW}$)
is positive at the center of bond, the QTAIM indication of ionic character.
%As before, $\tauKS$ is smooth and
%featureless, with an unusually severe minimum in the non-bonding side of Si.  
%
%
%As $\tauvW$ does not vanish in this marginally ionic
%bond, it is also interesting to 
It is also instructive to plot the electron
localization measure $\alpha$, shown as the lighter (cyan) solid line
in Fig.~\ref{fig:siofkden}.  In the SiO bond, this measure approaches
0.5, equivalent to an ELF of 0.67, which is the value approached by the 
other double bonds in the test set. A more telling structure 
occurs behind the Si atom, where $\alpha$ falls nearly to zero over an 
extended region of space.
The value $\alpha\!\sim\! 0$ (ELF of one) 
indicates that there is at most one occupied 
orbital so that $\tauKS$ reaches von~Weizs\"acker limit almost perfectly.
It also coincides with an abnormally low minimum in $\tauKS$.
This probably is a reflection of the hypovalent character of Si in this
molecule; however restricting the plot to the bond axis  
also eliminates the contribution of two $\pi$-bond orbitals 
to the KED.

\begin{figure}
\includegraphics[width=0.43\textwidth]{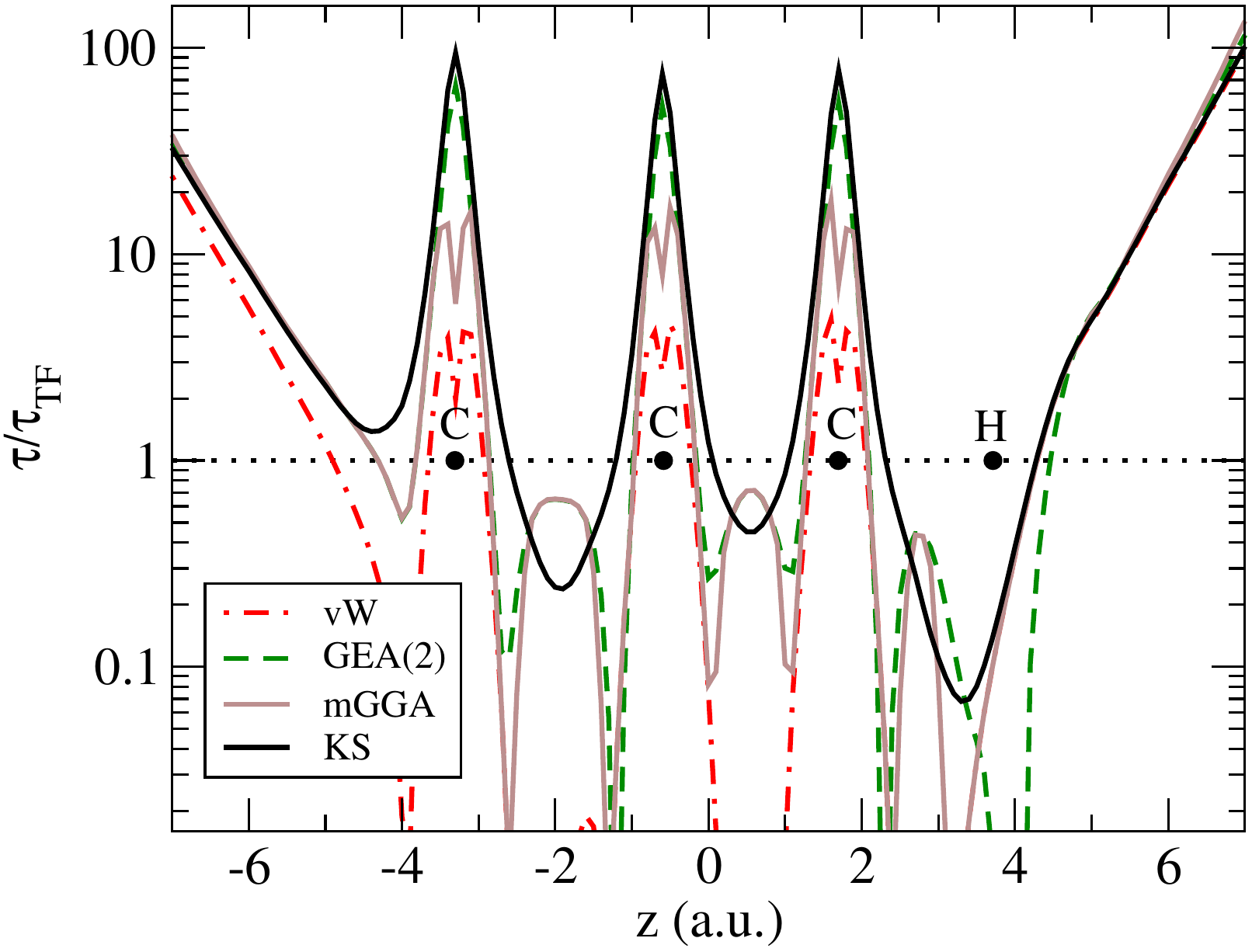} %.pdf}
\caption{\label{fig:C3H4fkden}
(color online)
The enhancement factor $F_S = \tau/\tauTF$, plotted 
on a log scale for various kinetic energy densities versus position along the 
carbon-carbon bond axis $z$ for the $\CthreeHfour$ (propyne) pseudo-molecule. 
Location of atoms on axis noted by a solid dot.
The Thomas-Fermi limit $F_S\!=\!1$ is shown as dotted line.
}
\end{figure}

Fig.~\ref{fig:C3H4fkden} shows enhancement factors
for the $\CthreeHfour$ pseudo-molecule, for points through the axis joining
the three C atoms and the on-axis terminal hydrogen (H$^4$). 
We plot $F_S$ on a log scale to focus on the 
situation at low densities, characterized by the carbon pseudo-atom cores
and the asymptotic region far from the molecule.

The asymptotic behavior of the Kohn-Sham and other KED's is 
dominated by linear trend of $\log{(F_S)}$ to infinity 
far from the molecule ($|z|\!>\!5$).
This is consistent with exponential decay of the charge density 
-- and with $\tauTF \sim n^{5/3}$ decaying more rapidly than any other 
model.  The three orbital-free models shown --
the von~Weizs\"acker model, the GEA and the mGGA -- have
roughly the same decay constant, and for the most part match up quite
well with the Kohn-Sham value.  Interestingly, the GEA is the best predictor
of $\tauKS$, performing better than the mGGA almost everywhere. 
The von~Weizs\"acker form almost matches the Kohn-Sham case for the 
asymptotic edge near the lone hydrogen ($z\!>\!5$) -- an indication that a 
single frontier orbital dominates the behavior of $\tauKS$ in this region.  
On the other edge of the bond axis ($z\!<\!-5$) $\tauKS$ is
roughly twice as large as $\tauvW$. This area sees the intersection of three 
frontier orbitals, one from 
each of the three C--H bonds that form tetrahedrally off the central bond axis.
This is enough to detach $\tauKS$ from the 
single occupied-orbital limit.

An interesting story also occurs in the pseudopotential cores of
the carbon atoms, with similar behavior seen for other atoms
that have had core electrons replaced by pseudopotentials.  
Although this
is arguably the least physical region of the molecule, 
it does represent one  %resemble the behavior of the KED in a 
of rapidly varying low density, but negligible density gradient, 
a topology that occurs
in noncovalent bonds and the interstitial regions of solids.  Here again the 
Kohn Sham KED is much larger than the Thomas Fermi value -- as noted before,
the charge and kinetic energy densities of our pseudopotential systems
observe a kind of complementarity, with one being large where the other is
small.  Of the three model KED's, it is the GEA that reproduces
the KS value most accurately.  The 
von~Weizs\"acker model peaks at the edge of 
the core region where $\gradnsq$ is large, disappearing in the 
center of the pseudopotential core where it goes to zero.  The GEA
here closely follows $\lapln$ which has a local maximum in the core
and thus the correct qualitative behavior; surprisingly, 
the result is even quantitatively accurate.
The mGGA trends more with $\tauvW$, and is severely deficient in magnitude.

\begin{figure}
\includegraphics[width=0.48\textwidth]{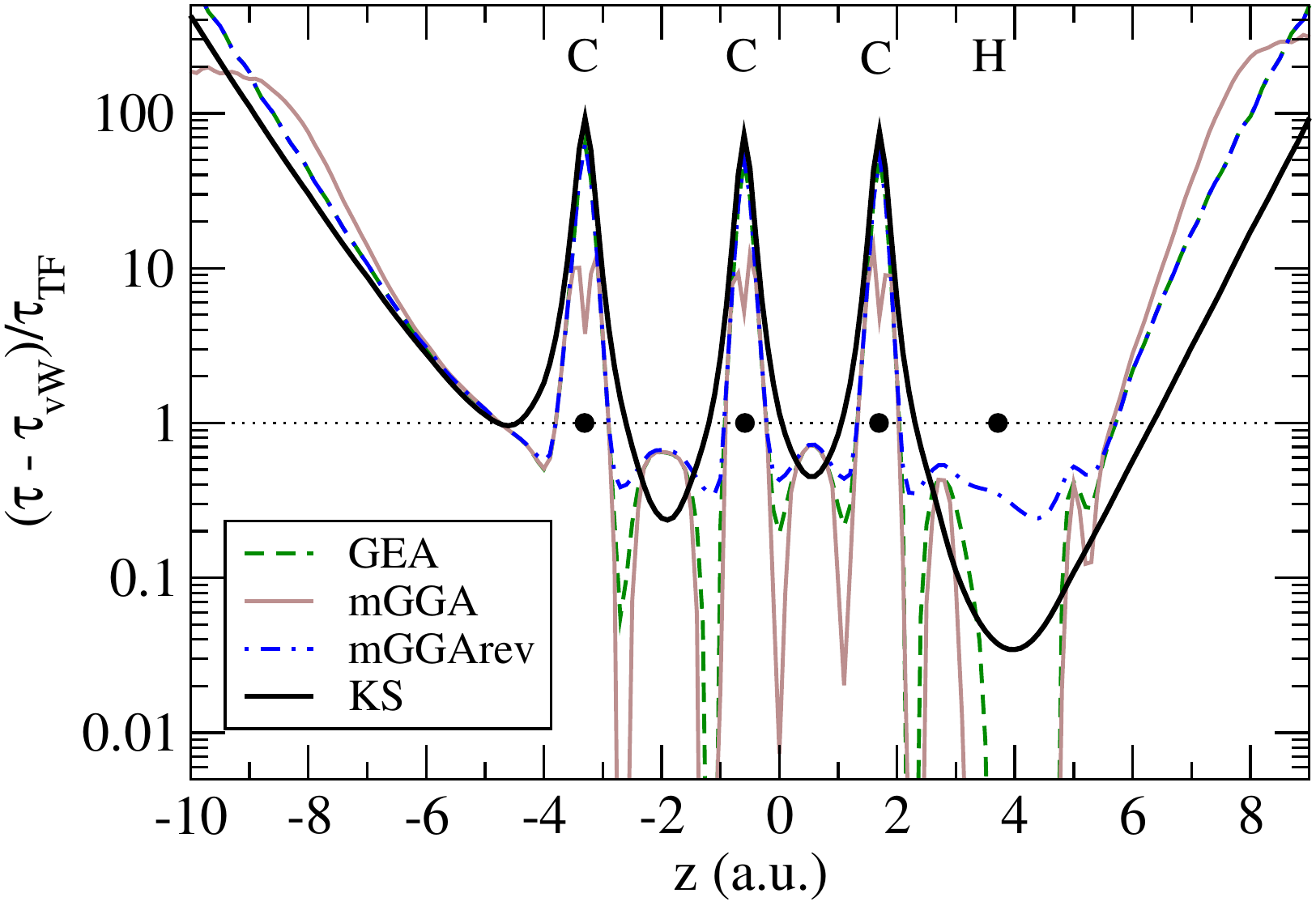}
%.pdf}
\caption{\label{fig:C3H4alpha}
(color online)
The electron localization measure $\alpha_{approx} = (\tau-\tauvW)/\tauTF$, 
plotted 
on a log scale for various kinetic energy densities versus position along the 
carbon-carbon bond axis $z$ for the $\CthreeHfour$ (propyne) pseudo-molecule.  
Location of atoms on axis noted by a solid dot.
The Thomas-Fermi value $\alpha\!=\!1$ is shown as dotted line. The
mGGArev is defined in Sec.~\ref{sec:discussion}.
}
\end{figure}

It is also useful for assessing approximate KED's to plot the 
approximation to the electron-localization factor $\alpha$ obtained
within a given model $\tau_{approx}$:  
  \be
     \alpha_{approx} = (\tau_{approx}-\tauvW)/\tauTF.
     \label{eq:alphaapprox}
  \ee
Focusing again on $\CthreeHfour$, which has the 
richest electronic structure of the test set, we plot 
in Fig.~\ref{fig:C3H4alpha} the $\alpha$ for several model KED's 
on a log scale versus position $z$ along the central bond-axis.
For the exact Kohn-Sham $\tau$, we find three regions with 
$\alpha\!<\!1$, an indication of electron localization -- the two carbon-carbon
bonds and single terminating hydrogen atom.
The other limit, $\alpha\!\gg\!1$, occurs inside the 
pseudo-atom cores and asymptotically.
The degree of localization of each small-$\alpha$ region
is consistent with the character of that region. It
is weakest ($\alpha\!\sim\!0.5$) for the triple bond between the 
second and third carbons, stronger ($\alpha\!\sim\!0.3$) for the single bond 
between the first two carbon atoms and extreme ($\alpha\!\sim\!0.05$) for the 
final hydrogen atom where only 
a single orbital is occupied.  As expected from the other figures shown,
no approximate model does very well in these important situations.

Two subtle order-of-limits issues come into play asymptotically. 
The approximate GEA and mGGA both do considerably
worse in predicting the asymptotic trend of $\alpha$ in 
Fig.~\ref{fig:C3H4alpha} than they do the 
enhancement factor of $\tauKS$, in~Fig.~\ref{fig:C3H4fkden}.  
$\alpha$ measures a difference 
between two models of $\tau$, and this difference is an order
of magnitude smaller than the value of either model far from the molecule.
It is thus a more
sensitive probe of error in orbital-free models. %of this quantity.
Secondly, while the GEA has the correct asymptotic behavior
(although consistently three times too large),
%far from the molecule.  
the mGGA has incorrect behavior as 
$|z| \rightarrow \infty$.  To understand this, note that $\alpha$
asymptotically can tend to anything from 0 to $\infty$.  
By the IP theorem, the numerator in
Eq.~(\ref{eq:alpha}) must vanish as the system tends to a one-electron state
and $\tauKS\!\rightarrow\!\tauvW$,
but the denominator also vanishes, as $n^{5/3}$, leaving the 
ratio undetermined.
While the asymptotic value of $\alpha$ is one in the mGGA,
for almost all the cases we have tested (SiO seems an exception) the observed 
limit is infinity.  This perhaps indicates only
that $\tauTF$ is an infinitely bad predictor of $\tau_{Pauli}$ for a region 
in which the Thomas-Fermi approximation fails.
%By the IP theorem, the Kohn-Sham
%KED tends to that of a single orbital and thus the von-Weizs\"{a}cker limit, 
%and thus, the difference in the numerator of $\alpha$ vanishes.
%However, the Thomas-Fermi KED vanishes exponentially faster than this 
%difference, so the ratio $\alpha$ diverges to infinity.  
%While the mGGA properly tends to 
%the von-Weizs\"{a}cker limit in this case, the secondary behavior -- how fast 
%the KED converges to the von-Weizs\"{a}cker limit, important
%for reproducing the ELF, is not captured by the mGGA, while it is by the GEA.
%ratio of this difference to the Thomas Fermi KED tends to infinity 
%-- the denominator decays exponentially faster to zero than the numerator.  
%Thus 
%The mGGA at large $z$ has incorrect asymptotic behavior,
%as it is designed to tend to the von-Weizs\"acker KED asymptotically, 
%and this is incorrect at least for practical values of $|z|$ here.

\section{Analysis\label{sec:discussion}}
\subsection{The GEA and asymptotic behavior of the KED}

It is worth analyzing in some depth
what happens in the region of asymptotic decay far from the molecule,
as demonstrated especially in Fig.~\ref{fig:C3H4fkden}, and to some degree
in Fig.~\ref{fig:S2fkden}.  
It is striking that $\tauKS$ and $\tauGEA$ 
match each other almost perfectly in this limit, within 3\%
at higher densities and no more than 15\% at the lowest densities we can
obtain. 
Consequently %the Laplacian of the density 
$\lapln$ and its GEA-level approximation, $\barlapln$, 
also agree almost exactly for this region.
This close agreement occurs for all systems studied, for example, for
SiO as one either moves away 
from the hypovalent Si atom or from the nearly filled O atom.
This is quite surprising since the 
GEA is designed for a completely different situation, that of the slowly 
varying electron gas, which is presumably unsuitable
for a classically forbidden region of space.
Formally, the regime of validity of the slowly varying electron gas is 
for systems for which the inhomogeneity measures $p$ and $q$ are 
everywhere $\ll 1$.  Obviously this criterion cannot be exactly met
for a molecule, but one might expect that, in any extended
region where these parameters are small,  $\tauKS$ should 
approach $\tauGEA$.  
In fact, the opposite proves true:
regions of space like covalent bonds, where $p$ and $q$ are consistently 
smallest,
are where the GEA does the worst, while the 
classically forbidden asymptotic region, where 
both $p$ and $q$ are much greater than one, is where it
performs best.
(To compare with the quantities shown in 
Figs.~\ref{fig:C3H4fkden} and~\ref{fig:C3H4alpha}, 
recall $\tauvW/\tauTF \!=\! 5p/3$ and $u_{vW}/\tauTF \!=\! 10q/3$.)   
Thus we have to conclude that some other 
phenomenon than the physics of the slowly-varying electron gas 
must explain the agreement asymptotically.

It is not hard to find one, at least qualitatively.
This region is characterized by an exponential decay of the 
density, $n\!\sim\!\exp{(-2kr)}$,
where $k = \sqrt{2I}$ gives the decay rate of the frontier 
orbitals, which have the highest eigenenergy, equal to the ionization 
potential $I$, and tunnel farthest into the vacuum.
As a result, the Kohn-Sham KED should behave as $k^2 n$, decaying
at a rate proportional to the local particle density. 
In contrast the Thomas-Fermi KED varies as $k_F^2 n$ with 
$k_F\!\sim\!n^{1/3}$.
The enhancement factor 
$F_S$ needed to correct $\tauTF$ to the Kohn-Sham value then 
scales as $k^2/k_F^2$, causing the
exponential growth seen in %in $F_S$ seen in 
Figs.~\ref{fig:S2fkden} and~\ref{fig:C3H4fkden}.
It is notable that the second-order gradient expansion reproduces 
this scaling behavior.  
The inhomogeneity variable $p$   %%%, proportional to $\gradnsq$,
is equal to $(k/k_F)^2$ for any exponentially decaying particle density
-- and $q$ is also, up to a %correction $\sim (k/k_F)^2 / kr$ due to curvature. 
correction due to curvature. 
The form of $\tauGEA$ [Eq.~(\ref{eq:taugea})]
gives its enhancement factor the correct limiting behavior 
as $r \!\rightarrow\! \infty$.  
In contrast, the fourth-order correction
has terms which scale like $p^2\!=\!k^4/k_F^4$,
which blow up exponentially as $r \!\rightarrow\! \infty$.  
And a GGA, a closed expression summing over all orders of
the gradient expansion, is not necessary to capture order-of-magnitude
trends and can 
actually be less accurate than the second-order GEA.

This is in stark contrast to what happens for the exchange energy: the 
energy density associated with a single frontier orbital behaves asymptotically
as $(\frac{1}{2r})n$ while the LDA scales as $k_F n$.  
Applying the second-order gradient
expansion to the LDA creates an exchange energy density that 
scales incorrectly as 
$(k^2/k_F)n$ and a potential that diverges 
exponentially.  A GGA is needed to produce an accurate exchange energy and a 
potential that %if not reproducing the correct $1/r$ asymptotic trend, 
is finite (if not with the correct $1/r$ form.)  This contrast between the correction
needed for LDA exchange $\sim n^{4/3}$ and Thomas-Fermi KED $\sim n^{5/3}$
contradicts the conjointness conjecture in its usual formulation -- 
the same form of enhancement factor cannot be optimal for both cases.

\subsection{Revisiting the mGGA}
It is not hard to diagnose why the mGGA KED has difficulty 
modeling the molecular bond. 
The mGGA was tested primarily for closed-shell atoms 
and several model one-dimensional systems.~\cite{PerdewConstantin}
The most serious defects of the mGGA seen in the current study are
associated with regions of %pseudopotential molecules that explore values 
of joint $\{p,q\}$ space that these systems do not access.  
The issue of vanishing KED 
is strongly correlated with regions where $p\!\sim\!0$ and $q$ is negative 
and of the order of unity, a combination that does not happen with 
atoms.~\cite{CancioExqlaplFull}  A second problem is the mGGA's
large underestimate in the pseudopotential core where $p\!\sim\!0$ and 
$q \!>>\! 0$; in atoms, a large $q\!>\!0$ is associated with finite and
normally large $p$, and occurs primarily
in the asymptotic region far from the atom where both tend 
to $\infty$.~\cite{loopdeloopfootnote}
While both these errors lead to underestimates of the KE,
the former is a failure to model the KED in the valence shell
of an atom or molecule and should have a large impact on the model's ability
to predict molecular structure.

\begin{figure}
\includegraphics[width=0.45\textwidth]{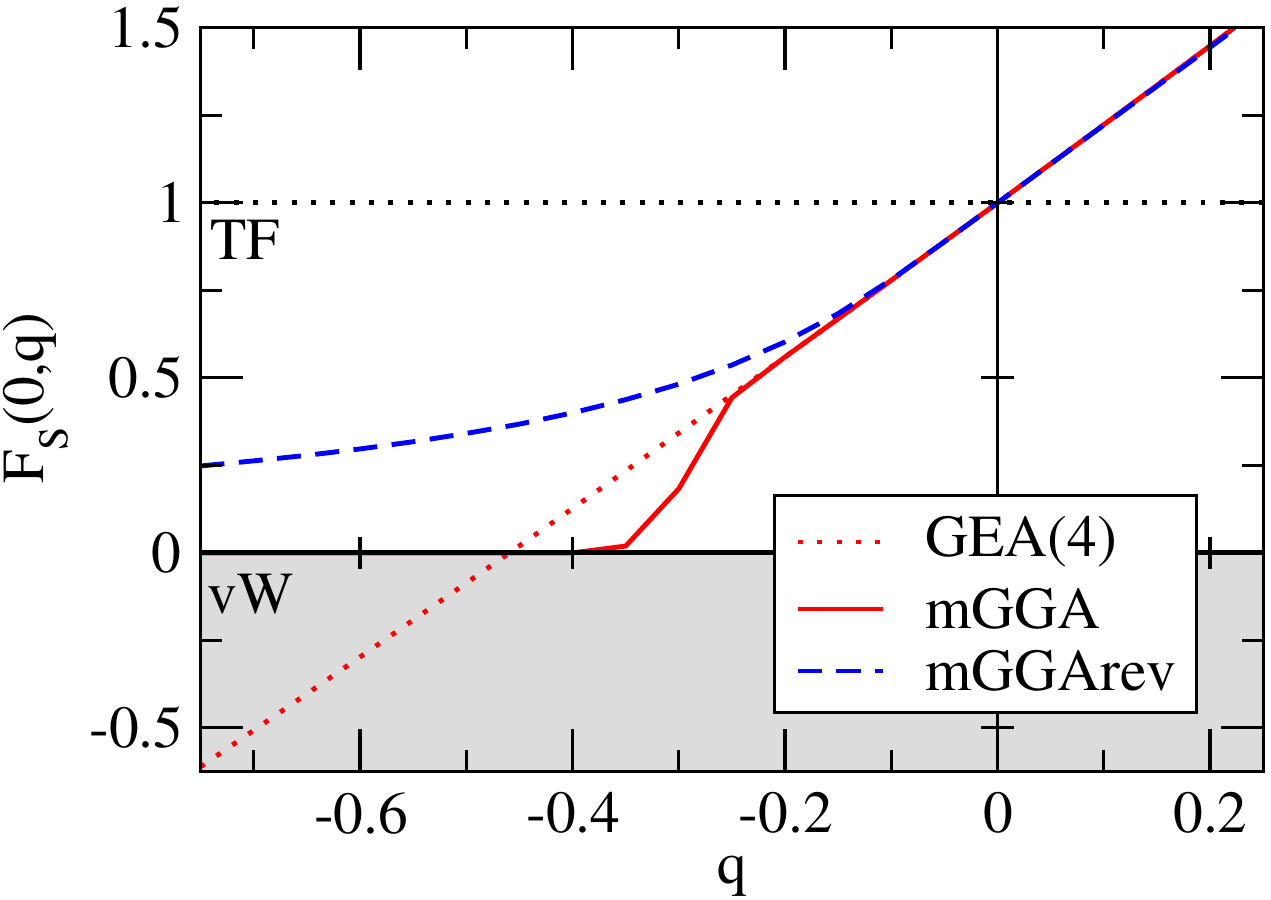} %.pdf}
\caption{
(color online)
Kinetic energy density enhancement factor $F_S(p,q)$ 
for orbital free kinetic energy density models,
plotted versus gradient-expansion parameter $q$ for $p\!=\!0$.
Thomas-Fermi limit $(F_S\!=\!1)$ indicated by dotted line.
Gray shaded region 
shows values of $F$ disallowed by the von~Weizs\"acker 
bound given by $F_S\!=\!0$.
\label{fig:fmggapzero}
}
\end{figure}

The common thread here is the behavior of the KED as a function of $q$ 
for values of $p \ll |q|$ and near to zero.
In Fig.~\ref{fig:fmggapzero} we plot the $p=0$ limit of the KED
enhancement factor, $F_S(0,q)$, for several KED models as a function of the 
scale-invariant factor $q$.
By definition, $F_S\!=\!1$ for $\tauTF$, %is simply one, 
as shown by the dotted horizontal line.  Likewise,
the von Weizs\"acker KED for $p\!=\!0$ is zero for all $q$.
The $F_S$ of the fourth-order gradient expansion 
approximation %(red dotted line) 
reduces to $1 + (20/9)q + (8/81)q^2$. 
This is nearly indistinguishable to the second-order gradient expansion, 
linear in $q$, because the fourth order coefficient is so small. 
The model of interest is the solid red line, that of Perdew and Constantin.  
It starts off with the gradient expansion and applies further 
constraints.  First, the von~Weizs\"acker bound requires that the 
enhancement factor be greater than $F_{vW}$, in effect greater than zero. 
For $q\!<\!0$ the KED must 
transition fairly quickly from GEA-like behavior to zero, as the GEA
breaks this constraint at $q\!\sim\!-0.45$.
The second imposed limit is that the enhancement factor goes to 
$1 + F_{vW}$ in the limit of large positive $q$, seen for example
in our data in pseudo-atom cores, but not shown in 
Fig.~\ref{fig:fmggapzero}.

The flaws in the mGGA seem to be caused by its 
implementation of these constraints.
The most important, the extinction of KED for negative $q$ and
small $p$, is clearly the result of a transition 
from $\tauGEA$ to $\tauvW$ that zeroes out the KED for 
$q\!<\!-0.30$, a value of $q$ achievable in the vicinity of an atomic
lone-pair or a covalent bond, especially in a pseudopotential system.
This transition scheme implicitly
invokes an order of bounds as follows:
\be
    \tauGEA \ge \tauKS \ge \tauvW
    \label{eq:theirvwconstraint}
\ee
That is, it seeks an interpolation \textit{between} the two limiting cases,
which leaves very little room for smoothing out 
the transition.~\cite{Wigglefootnote}
It makes more sense to try an interpolation \textit{above} the two
limiting cases, assuming a constraint
\be
    \tauKS \ge \mathrm{max}(\tauGEA,\tauvW),
    \label{eq:myvwconstraint}
\ee
demonstrated by the blue dashed curve in Fig.~\ref{fig:fmggapzero}.
Such a transition is smoother and thus more physically
appealing, and has the effect of enhancing rather than reducing 
the KED in high-density, low-$q$, regions.  
A smoother transition to zero should also produce a smoother
kinetic energy potential 
which is important for a self-consistent density
functional minimization.  Abrupt changes in the enhancement factor of a 
Laplacian-based density functional 
can be disastrous when taking functional
derivatives with respect to $\lapln$, since these involve
derivatives of the density up to $\nabla^4 n$.~\cite{CancioExlapl}

The mGGA model can also be improved by relaxing the large-$q$ cutoff
that it imposes.  
As seen especially in 
Fig.~\ref{fig:C3H4alpha}, the mGGA clearly overcorrects 
for the regions where $q\!\gg\!1$, %\rightarrow +\infty$,
in the pseudo-atom core and asymptotically.  
In both cases, the second-order GEA is a better approximation to
$\tauKS$ and there is less motivation for 
a GGA correction to it than is the case for exchange.

\subsection{\label{sec:mggarev} 
Revision of the mGGA and application to atomization energy}

We propose to make a revised mGGA following two simple points:
imposing the von~Weizs\"acker lower bound by means of 
Eq.~(\ref{eq:myvwconstraint}) and relying on the second-order
gradient expansion otherwise.  
This satisfies the constraints for the two main limiting cases of the KED -- 
that of delocalized electrons with slowly-varying density
and that of strong electron localization, and otherwise keeps physically 
reasonable behavior for regions of high inhomogeneity --  pseudo-atom cores and
asymptotic decay.
We first define a measure of electron localization $z$ that depends upon the 
difference between GEA and vW enhancement factors for the KED:
\be
    z = F_{GEA} - F_{vW} - 1 = \frac{20}{9}q - \frac{40}{27}p
\ee
The factor $z$ can be thought of as a poor-man's electron localization
factor -- an orbital-free expression for the $\alpha$ used to
describe electron localization in metaGGA's and from which the ELF is 
constructed.  
We then look for the simplest possible 
asymptotic transition between $F_{GEA}$ and $F_{vW}$ that 
imposes the von~Weizs\"acker bound, which occurs for $z\!\leq\!-1$. 
Adapting a form recently used to construct a $\lapln$-based exchange
function~\cite{CancioExqlaplFull} results in an enhancement factor 
\be
      F^{mGGArev}_S = F_{vW} + 1 + z\left\{1 - \exp{\left(\frac{1}{z}\right)}\left[1-H(z)\right] \right\}
      \label{eq:mggarev}
\ee
where $H$ is the Heaviside step function.  
This is shown in Fig.~\ref{fig:fmggapzero} as a function of $q$ for $p\equal 0$.

The limiting behavior of this correction can be characterized by three cases,
roughly analogous to those defined by the ELF:
\benum
    \item If $z \rightarrow 0$, then both $p$ and $q$ must become small.
    The density is slowly varying, and close to the homogeneous gas limit,
    typical of metallic bonds.
    In this case, $F_S$ goes to the gradient expansion form:
    \be 
         F_S\!\sim\!F_{vW} + 1 + z = F_{GEA}
    \ee
   
    \item If $z\!<\!0$, this means that either $q$ becomes negative or 
    $p \rightarrow +\infty$ with a finite $q$.  
    In this case, $F_S$ approaches the von~Weizs\"acker limit:
    \be
        F_S \rightarrow F_{vW} + O(1/z).
    \ee
    This is the proper description of a region with strong electron 
    localization, such as a covalent bond.

    %the remainder of the KED, which
    %contributes to the nonsingular of  
    %The former case may be seen in any atom, near the nucleus (inside
    %the 1s shell).  Here the wave function must have a cusp in it since the
    %minus the Laplacian of the wave function must go as $+Z/r$ to cancel the 
    %effects of a $-Z/r$ potential in Schrodinger's equation.  This guarantees
    %that $\lapln$ will have a $-Z/r$ singularity as $r \rightarrow 0$
    %and thus $q$ also.  Large but not infinite values of $p$ at finite
    %$q$ are a characteristic of the charge density in the region between
    %two atomic shells, such as the outer core shell and valence shell
    %of a light atom.  In either case, the limit we take here is 
    %Of particu
    %\bea
    %    F_S &\!\sim\!& 1 + F_{vW} + z \left[ 1 - 1 - 1/z - 1/2z^2 + . . .\right] \\
    %        & = F_{vW} + O(1/z).
    %\eea
    %This result is basically correct for the nuclear cusp, if we assume only
    %one 1s shell participates,
    %and is Perdew and Constantin's basic
    %assumption for $p \rightarrow \infty$. 
 
    \item If $z\!\gg\!0$, we get the same result as for 
    $z$ small:
    \be
       F_S \rightarrow F_{GEA}.
    \ee
    The primary situation for which this limit applies is an 
    exponentially decaying density, for which 
    $p\!\sim\!q \rightarrow \infty$ and $z \rightarrow 20 q/27$.
    \eenum
    The final case also describes a situation with $q\!\gg\! 1$ and 
    finite $p$, seen here in pseudo-atom cores,  
    and in the transition between atomic shells in real atoms.  
   %For all regions with $z\!\gg\!0$
   %we have the semiempirical observation that the second-order GEA %for the KED
   %is well-behaved and even a fairly good approximation to $\tau$.

    Also of interest for real atoms is the limit 
    $q, z \!\rightarrow\! -\infty$  which 
    occurs near the nucleus and is caused by the cusp in the electron density.
    The functional derivative $\delta \tau / \delta n(\bfr)$,
    used for the self-consistent determination of the charge density
    in OFDFT, 
    must tend to $Z/r$ near the nucleus so as to cancel the $-Z/r$ 
    contribution from the electron-nucleus potential.  This behavior is 
    exactly given by the functional derivative 
    of $\tauvW$, and thus by $\tau_{mGGArev}$ as well. 
    The leading correction to $\tauvW$ is of order $1/q$; 
    %, and is vanishingly small in this limit; 
    its functional derivative is known to be finite,~\cite{UmrigarGonze} 
    but it can cause a sizable error in the cusp of $\tau$ at the nucleus.

To evaluate the effects of this revised mGGA, we plot its
enhancement factor for SiO in Fig.~\ref{fig:siofkden} and its approximation 
to $\alpha$ using Eq.~(\ref{eq:alphaapprox}) for propyne in
Fig.~\ref{fig:C3H4alpha}.  
As shown in the latter,
the mGGArev by construction follows the
GEA curve almost everywhere in space -- except for regions of electron
localization, where
it enhances the magnitude of the KED considerably over the 
GEA.  It is thus an improvement over both GEA and mGGA.
However the mGGArev overcorrects for situations of strongest electron
localization.  For the single C$^1$-C$^2$ bond of propyne, 
with a small $\alpha$ of 0.3, the mGGArev gives
a modest average overcorrection. 
It severely overcorrects for the most localized situations, 
where $\alpha \!<\!0.1$: near the terminal
H$^4$ atom in propyne and behind the hypovalent Si atom in SiO.
This problem may be ameliorated by 
tinkering with the rate of transition between GEA and vW limits in 
Eq.~(\ref{eq:mggarev}) -- in the current form (Fig.~\ref{fig:fmggapzero}),
it is probably too slow.
One region that shows little
change from the mGGA is behind the C$^1$ atom ($z\!\sim\!-4$) in 
Fig.~\ref{fig:C3H4alpha}.  This is not a region of electron
localization since it feels the overlap of three 
neighboring C--H bond orbitals
so the model has no criterion to correct for the error of the GEA.

In Table~\ref{table:keintegral} we show errors with respect to the 
integrated Kohn-Sham KE averaged over the test set, as a measure
of the overall quality of the models discussed in this paper.
A net trend across all models is the underestimation of 
the KE by roughly 10\%.  Unfortunately, by the virial theorem,
the total KE is equal in magnitude to the total energy,
which varies from 3.5~hartree for the valence shell 
of $\SiHfour$ to 31~hartree for that of $\CtwoHtwoOtwo$. 
Absolute errors in KE can thus be as large as several hartrees.  
While the second-order GEA is a modest improvement
over the Thomas-Fermi result, the
mGGA, in attempting to address the limitations of the GEA, actually loses 
some of the ground gained by it.  
The revised mGGA introduced here
is more consistently an improvement.  One situation
in which it is not, $\SiHfour$, results in the maximum RE being three
times the MARE, and an overestimate, not an underestimate.  
As shown in Fig.~\ref{fig:SiH4ppr}, this molecule is 
marked by a substantial region that is 
near the von~Weizs\"acker limit, $\alpha\!\sim\!0$, 
for which the mGGArev overestimates the KED.  This again indicates a need for 
further exploration of how to manage the transition from delocalized to 
localized electronic systems.  % could be of value.

\begin{table}[t]
\begin{tabular}{r|c|c|c|c|c}
\hline\hline
                & TF     & GEA & mGGA & mGGArev & VT84F \\
\hline
MRE             & -0.162  & -0.112  & -0.124  & 0.0021 & 0.229 \\
MARE            &  0.162  &  0.112  &  0.139  & 0.0873 & 0.229 \\
Max RE          & -0.202  & -0.159  & -0.178  & 0.233 & 0.550 \\
\hline\hline
\end{tabular}
\caption{\label{table:keintegral} 
Mean relative error (MRE), absolute relative error (MARE) and maximum
relative error (Max RE) for  various orbital-free estimates of the 
Kohn-Sham kinetic energy. 
}
\end{table}

To further characterize the quality of our revised mGGA, we calculate 
the atomization energies of the AE6 test set.  This 
helps gauge the extent to which systematic errors 
in the total energy are cancelled out in taking energy differences.
This is done not self-consistently, using conventional PBE Kohn-Sham densities  
and bond lengths (Table~\ref{table:bond length}).
The results are shown in Table~\ref{table:orbitalfreeAE}.
\begin{table}
\begin{tabular}{c|r|r|r|r|r|r|r}
\hline \hline
System & Exp. & KS  & VT84F &  mGGA- & mGGA & GEA & TF\\
&  & & &  rev & & & \\
\hline
$\SiHfour$      & 322.4 & 315.9 & -178.1 & -14.9 & 57.2 & -183.8 & -174.9\\
$\Stwo$         & 101.7 & 124.4 & 140.9 & 17.7 & -101.5 & -72.1 & -100.6\\
SiO             & 192.1 & 205.4 & -4.8 & -4.5 & -169.3 & -97.6 & -213.7\\
$\CtwoHtwoOtwo$ & 633.4 & 680.6 & 476.6 & 240.2 & -422.8 & -119.0 & -416.6\\
$\CthreeHfour$  & 704.8 & 726.6 & 581.9 & 572.3 & 24.2 & 115.9 & 35.7\\
$\CfourHeight$  & 1149 & 1175.3 & 1072.4 & 811.8 & 142.6 & 96.0 & -53.3\\
\hline
MAE            & --  & 23.0 & 182.1 & 246.8 & 595.5 & 560.7 & 671.1\\
%\hline
%$\SiHfour$      & 322.4 & 315.88 & -178.05 & -14.87 & 57.22 & -183.78 & -174.87\\
%$\Stwo$         & 101.7 & 124.41 & 140.94 & 17.67 & -101.48 & -72.13 & -100.64\\
%SiO             & 192.1 & 205.35 & -4.81 & -4.53 & -169.28 & -97.63 & -213.65\\
%$\CtwoHtwoOtwo$ & 633.4 & 680.61 & 476.60 & 240.15 & -422.83 & -118.95 & -416.63\\
%$\CthreeHfour$  & 704.8 & 726.56 & 581.93 & 572.31 & 24.19 & 115.90 & 35.71\\
%$\CfourHeight$  & 1149 & 1175.34 & 1072.42 & 811.78 & 142.59 & 95.95 & -53.34\\
%\hline
%MAE            & --  & 22.97 & 182.14 & 246.82 & 595.50 & 560.67 & 671.14\\
%% Old -- some not at molecular minima, rounding conditions on metaKDEN sucked.
%$\SiHfour$      & 322.4 & 315.88  & -15.77 & 57.22   & -183.78 & -174.88\\
%$\Stwo$         & 101.7 & 124.41  & 18.06  & -101.48 & -72.13 & -100.64\\
%SiO             & 192.1 & 205.35  & -4.09  & -169.28 & -97.62  & -213.65\\
%$\CtwoHtwoOtwo$ & 633.4 & 680.61  & 241.27 & -422.83 & -118.96 & -416.64\\
%$\CthreeHfour$  & 704.8 & 726.56  & 572.33 & 24.18   & 115.89  & 35.71\\
%$\CfourHeight$  & 1149  & 1175.34 & 811.31 & 142.54  & 95.62   & -53.59\\
%\hline
%MAE & -- & 23.00  & 246.71  & 595.51  & 560.73  & 671.18\\
\hline \hline
\end{tabular}
\caption{
Atomization energies for the AE6 test set in kcal/mol.  Shown are 
experimental values from Ref.~\onlinecite{AE6}, 
self-consistent Kohn-Sham results, 
and results of orbital-free models evaluated with the Kohn-Sham density. 
%Thomas-Fermi (TF), second-order gradient expansion (GEA), 
%Perdew-Constantin mGGA and
%our revision (mGGArev).
Also shown is the mean absolute error with respect to experiment.
\label{table:orbitalfreeAE}
}
\end{table}
First we note how far the Thomas-Fermi atomization
energy is from experiment, with an MAE ten times worse than the 
LDA and over thirty times worse than the PBE Kohn-Sham models.  
It almost always fails to predict binding, at best 
giving a marginal binding energy.  
The second-order GEA does provide a modest
improvement over the TF case, but again shows severe under-binding.
By respecting the von~Weizs\"acker lower bound, the mGGA ought to 
significantly improve GEA atomization energies.  % over those of the GEA.
Instead it performs worse for the majority of the test set, and 
in some cases worse than Thomas-Fermi.
In constrast, the mGGArev does show the
expected improvement over Thomas-Fermi and GEA.
It binds all but one molecule, $\SiHfour$, the standout worst case in
Table~\ref{table:keintegral}, and the one case that the mGGA binds. 
On average it removes 60\% of the AE error of the TF and 
for one or two systems almost approaches the LDA in quality.
However its MAE is still an order of magnitude worse than the PBE and
a factor of three worse than the LDA.

To put these results in perspective, we perform calculations
for the VT84F,~\cite{KCST13} a nonempirical GGA for the kinetic energy.  This applies
the key constraints of the mGGA -- respecting the gradient expansion
in the small-$p$ limit and requiring the von Weizs\"{a}cker constraint
for all $p$; in addition it enforces the non-negativity~\cite{LevyHui} 
of the Pauli potential, $\delta \tau_{Pauli}/\delta n(\bfr) \ge 0$.  
The VT84F total kinetic energy (Table~\ref{table:keintegral}) 
is by a large margin the least accurate
of all models considered, including the Thomas-Fermi model.
However it has the overall best prediction of atomization energies 
(Table~\ref{table:orbitalfreeAE}), and 
fails significantly only for $\SiHfour$.  It may be hard to enforce 
both the GEA and the constraint $\tau > \tauvW$ with only access to $\gradnsq$
as a variable and not 
overestimate the total KE. However enforcing constraints on the potential -- an 
infinitesimal energy difference -- seems to help for predicting accurate
finite-energy differences.  It is reassuring that the simple metaGGA 
we present here is comparable in quality to the VT84F without (as yet) 
taking the potential into consideration. 

\section{Discussion and Conclusions\label{sec:conclusion}}
We present highly converged DFT calculations 
for the AE6 test set of molecules, within a plane-wave pseudopotential
approach.  
We use these to visualize the Kohn-Sham kinetic energy density and
related quantities that are ingredients of modern DFT's, specifically metaGGA
models for the exchange-correlation energy, and orbital-free models for 
the KED.  
By providing a highly accurate
map between density and kinetic energy density for physically reasonable
model systems, our data enables the use of visualization techniques
employed in the qualitative analysis of  electronic structure 
to test approximations to this critical ingredient for DFT.
The pseudopotential method works especially well in characterizing
the classically forbidden region far from nuclei, and is reasonable in
its description of bonds; its main limitation is the loss of knowledge
of the core region, most importantly, the character of the core-valence
transition that plays a key role in determining bond lengths.

The choice of the AE6 test set does not break new ground in 
visualization of electronic structure, but does an excellent job of 
illustrating many of the lessons learned from QTAIM and other visualization 
approaches, 
particurly the role of $\lapln$ in understanding valence electronic structure 
and the KED in measuring electron localization.  
The SiO molecule is perhaps of most interest structurally,
given the relationship between the hypovalent character of Si and the
strong indication of electron localization in the Si valence shell; 
also of interest is the identification of the bond as ionic rather 
than polar covalent by QTAIM criteria.  

A major finding of the paper is the surprising success of the 
gradient expansion expression for the Kohn-Sham KED.  The gradient
expansion approximation $\lapln\!\sim\!6(\tauKS \!-\!\tauTF)$ used
in modern metaGGA's is at least qualitatively
very good -- $\lapln$ to some degree picks up the complementary behavior
of the kinetic and particle densities, and detects regions where one is
larger than the other.  Rather surprisingly, this approximation is 
the most accurate in the lowest density regions, in the classically
forbidden regions far from nuclei.  
This is because it has the exact asymptotic behavior with respect to distance 
from nuclei and not too bad quantitative values for all systems considered.  

The asymptotic exactness of the GEA, although not news, is worthy of note 
since it points out the limitations of the idea of conjointness between
exchange-correlation and kinetic energy functionals, both as a conjecture
and as a design philosophy.  Lessons learned in designing functionals 
for the former case do not necessarily transfer over to the latter.
The very different behavior of the 
gradient expansion for exchange in the asymptotic limit necessitates a 
fundamentally
different functional form for exchange energy GGA's and kinetic energy GGA's.  
The gradient expansion of the former must be controlled by some form of cutoff 
at large values of reduced density gradient $p$
while that of the latter, as best we can see, is better off mostly untouched.

A second point underscores the difficulty in building orbital-free
models of the KE -- the gradient expansion behaves 
worst in describing ``slowly varying" regions of space -- where the 
inhomogeneity parameters $p$ and $q$ used to describe it are small.
For the KE density, there seems not to be a good ``semilocal" approximation
for real systems -- one cannot rely on $p$ and $q$ being small locally
to predict that the gradient expansion should hold locally, in contrast
with the XC energy density.  %There is a 
When taken separately, exchange and correlation energy
densities have similar problems to those we see here for the 
KED; however, there is a notable cancellation of error
between the two that makes semilocal approximations work better
than might be expected.~\cite{Holefootnote} 
%(In any high density region, the exchange hole around any 
%electron due to Pauli statistics tends already to be mostly centered around that 
%electron's position and similar in shape and size to the corresponding hole
%in the local density approximation; to the extent it is not, it generates a 
%strong nondynamic correlation response that tends to restore the overall XC hole 
%back to the isotropic case.)
What the KED lacks then is a companion mechanism such as correlation by 
which deviations from the GEA can be cancelled out.
This failure does not contradict the idea of the gradient expansion.  
The limit in which it is exact is that of \textit{globally} small $p$ and $q$, 
with the result of delocalized electronic orbitals almost everywhere,
a condition that is not met by any molecule.
%It is 
%error cancellation between exchange and correlation that make semilocal
%approximations work for the XC energy for small systems,
%rather than the appropriateness of the gradient expansion.

The other major finding of this paper relates to 
the Perdew-Constantin mGGA model of the kinetic energy density.  
We have found a number
 of problems which degrade its 
performance with respect to the Thomas-Fermi model. 
Its description of the KED in regions of high inhomogeneity and
low density are less effective than the simpler second-order GEA.
More importantly, it is subject to an ``extinction"
effect for large negative values of the reduced Laplacian $q$, 
causing it to plummet to zero in regions of covalent bonding.  This effect 
is caused by the particular form used in the imposition of the 
constraint $\tau\!>\!\tauvW$, which becomes important in regions of 
electron localization, such as covalent bonds.  
It is aggravated by the use of pseudopotentials, which exaggerate
the magnitude of negative-$\lapln$ or VSCC regions in comparison to their
all-electron counterparts.  
The correlation between this effect and electron localization seems
responsible for the poor binding seen with this model.  The formation of bonds 
can reduce electron localization and thus reduce the extinction effect 
relative to the isolated atom case, 
leading to a lack of error cancellation in taking energy differences. 
Notably, the cases in which the mGGA gives improved binding energies,
$\SiHfour$ and $\CfourHeight$, are the ones with exclusively single 
bonds and thus roughly the same degree of electron localization in molecule 
and atom.

This work points to several avenues of future research. 
The mGGArev form we propose for the KED %has room for improvement. 
is the simplest, not best, form that can fit the constraints imposed in
Sec.~\ref{sec:discussion} and should perhaps be used not as a finished 
functional but as an indication of how to proceed in developing one.
Particularly, the use of an ``orbital-free ELF", using derivatives
of the density to approximate the ELF and its ability to distinguish
between different kinds of bonds, seems worthy of further investigation.
However, in its current form, our model regresses on the mGGA's capacity to 
handle covalently bonded hydrogen
atoms and other situations of nearly perfectly localized electrons.  
Notably, our proposed 
constraint, that $\tau\!>\!\tauGEA$ when $\tau\!\rightarrow\!\tauvW$, is not 
universal -- it fails for the 1s shell of atoms, as shown
in Ref.~\onlinecite{PerdewConstantin}.
Not surprisingly, the mGGA functional, with $\tau\!<\!\tauGEA$ 
in this limit was arrived at partly through the consideration of this case.  
However our constraint %%Eq.~(\ref{eq:myvwconstraint}) 
does appear to be valid for any other shell of an all-electron atom 
-- and it is responsible for our current revision's relative success in
predicting binding energies of the AE6 test set. 
Any more sophisticated
model will thus have to ameliorate somehow the problems for hydrogen encountered
by the mGGArev while keeping its nice features for bonding.

A second notable issue is the large deviation of 
$\lapln$ and $\tauKS$ obtained with pseudopotentials
from their all-electron values just inside the pseudopotential
cutoff radius.  
As noted earlier,  
the resulting exaggeration of the negative value of $\lapln$
in VSCC's contributes to the failure of the mGGA to predict binding in 
pseudopotential systems.  
But, given the sudden switching behavior that
the mGGA shows for negative $q$ (Fig.~\ref{fig:fmggapzero}), 
it is quite possible that with the smaller $q$
values of all-electron systems, this effect would not be an issue.
This raises the 
question of how much pseudopotentials that have been constructed
to match valence density alone can be trusted in metaGGA or OFDFT 
applications that rely upon variables that are more sensitive to changes in 
electronic structure.

In the big picture, the ability to get orbital-free DFT's that are competitive 
with Kohn-Sham approaches remains a challenge.  However, some progress towards 
functionals useful in extreme situations where Kohn-Sham
approaches are impractical may yet be done with metaGGA's working
with semilocal properties of the density.  
Visualization can be a useful tool
in this process, fruitfully bringing together strands of qualitative
and quantitative thinking about electronic structure.

\begin{acknowledgments} A.C.C would like to thank John Perdew 
and Paul Grabowski for useful discussions.  Work supported in
part by National Science Foundation grant DMR-0812195.
\end{acknowledgments}

%\bibliography{exc_fit_new,DFT,ked,ELF,footnote}     

\begin{thebibliography}{92}
\bibitem [{KT1()}]{KT15footnote}%
  \BibitemOpen
  \href@noop {} {}\bibinfo {note} {However we note it may not be possible
  fundamentally to build approximate KE functionals that work well for both
  exact and pseudopotential situations -- see
  Ref.~\onlinecite{KT15}}\BibitemShut {NoStop}%
\bibitem [{loo()}]{loopdeloopfootnote}%
  \BibitemOpen
  \href@noop {} {}\bibinfo {note} {For $Z\!<\!10$, large $q$ and $p$ can occur
  in the transitional region between shells,~\cite{CancioExqlaplFull} but for
  $Z \!\geq\! 10$ these effects diminish to zero as the cores of larger-$Z$
  atoms approximate the Thomas-Fermi limit.}\BibitemShut {Stop}%
\bibitem [{Wig()}]{Wigglefootnote}%
  \BibitemOpen
  \href@noop {} {}\bibinfo {note} {This effect can be seen in Fig.~2 of
  Ref.~\onlinecite{PerdewConstantin} as an unphysical wiggle in the enhancement
  factor $F_s(s,q)$ plotted versus $s = \sqrt{p}$ for a few selected values of
  $q$. Significantly, the one choice of $q\!<\!0$ shown in the original paper
  was $q \!=\! -10$, so that the behavior at small negative $q$ could not be
  detected.}\BibitemShut {Stop}%
\bibitem [{Hol()}]{Holefootnote}%
  \BibitemOpen
  \href@noop {} {}\bibinfo {note} {This cancellation arises because of the
  interrelationship between exchange and correlation in defining the XC hole
  about any electron. Any nonlocal feature in the exchange hole generates
  nonlocal, nondynamic correlations which tend to screen out its
  effect.~\cite{CancioChouHood} No such effect is known for the
  KED.}\BibitemShut {Stop}%
\end{thebibliography}

%\end{document}

\end{document}